\documentclass[a4paper,11pt]{article}\pdfoutput=1
\usepackage{subcaption}

\usepackage{jinstpub}
\usepackage{adjustbox}
\usepackage{multirow}
\usepackage{graphicx}
\usepackage{tabularx} 
\usepackage{booktabs}  
\title{Development and first performance evaluation of multi-element monolithic HPGe detector for X-ray spectroscopy}

\author[a,1]{N.~Goyal\note{Corresponding author}}
\author[a,2]{F.J.~Iguaz}
\author[k]{S.~Aplin}
\author[b]{A.~Balerna}
\author[c]{P.~Bell}
\author[l]{J.~Casas}
\author[c]{M.~Cascella}
\author[d]{S.~Chatterji}
\author[e]{C.~Cohen}
\author[e]{E.~Collet}
\author[d]{E.N.~Gimenez}
\author[f]{H.~Graafsma}
\author[f]{H.~Hirsemann}
\author[c]{K.~Klementiev}
\author[g]{T.~Kolodziej}
\author[e]{T.~Martin}
\author[h,i,j]{R.H.~Menk}
\author[a]{C.Menneglier}
\author[e]{C.~Meraihia}
\author[d]{J.R.~Murias}
\author[k,n]{M.~Porro}
\author[l]{M.~Quispe}
\author[m]{B.~Schmitt}
\author[d]{S.~Scully}
\author[k]{M.~Turcato}
\author[c]{C.~Ward}
\author[f]{E.~Welter}

\affiliation[a]{SOLEIL Synchrotron, L'Orme des Merisiers, D\'epartementale 128, 91190 Saint-Aubin, France}
\affiliation[b]{INFN, Frascati National Laboratory, Via Enrico Fermi 54, 00044 Frascati, Italy}
\affiliation[c]{MAX IV Laboratory, Lund University, Fotongatan 2, Lund 224 84, Sweden}
\affiliation[d]{Diamond Light Source Ltd, Harwell Science and Innovation Campus, Didcot OX1 10DE, United Kingdom}
\affiliation[e]{European Synchrotron Radiation Facility (ESRF), 71 avenue des Martyrs, 38043 Grenoble, France}
\affiliation[f]{Deutsches Elektronen-Synchrotron DESY, Notkestr. 85, Hamburg 22607, Germany}
\affiliation[g]{Jagiellonian University, ul. Golebia 24, Krak´ow 31-007, Poland}
\affiliation[h]{Elettra Sincrotrone Trieste S.C.p.A., Trieste 34149, Italy}
\affiliation[i]{INFN Trieste, Trieste, Italy}
\affiliation[j]{Department of Computer and Electrical Engineering, Mid Sweden-University, Sweden}
\affiliation[k]{European XFEL, Holzkoppel 4, Schenefeld 22869, Germany}
\affiliation[l]{ALBA-CELLS Synchrotron Radiation Facility, Carrer de la Llum 2-26, 08290 Cerdanyola del Valles, Spain}
\affiliation[m]{Paul Scherrer Institute, Forschungsstr. 111, Villigen 5232, Switzerland}
\affiliation[n]{Department of Molecular Sciences and Nanosystems, Ca’ Foscari University of Venice, 30172 Venezia, Italy.”}
\emailAdd{$^{1}$nishu.goyal@synchrotron-soleil.fr}
\emailAdd{$^{2}$francisco-jose.iguaz-gutierrez@synchrotron-soleil.fr}

\abstract{The first operational prototype of a high-purity Germanium (HPGe) detector developed within the European LEAPS-INNOV project is presented in this work. This prototype features a monolithic, multi-element sensor optimized for high-resolution X-ray spectroscopy in the hard X-ray regime, capable of handling high count rates (20-250~kcps/mm$^{2}$) across a broad energy range (5-100 keV). We discuss here a complete laboratory-based characterization of the detector's performance, as well as an on-beam evaluation at the BM05 beamline of the ESRF synchrotron facility, using monochromatic X-rays in the 20–50~keV energy range. We provide a detailed performance assessment that also includes a phenomenological defect-depth estimation model.}

\begin{document}
\maketitle
\section{Physics Motivation}
\label{sec:physics motivation}
The pursuit of advanced spectroscopic techniques at upgraded synchrotron radiation facilities demands detectors with high quantum efficiency and spectral resolution, especially in the hard X-ray energy range (above 10 keV). High-purity Germanium (HPGe) detectors offer significant physical advantages over Silicon-based detectors in this context, making them a better choice for techniques such as X-ray Absorption Fine Structure (XAFS) and X-ray fluorescence (XRF)~\cite{calvin2013xafs,Hubbard1996}.\\
Fundamentally, the choice of Germanium over Silicon is motivated by its higher atomic number (Z=32) vs. (Z=14) for Si, which enhances the photoelectric absorption cross-section, particularly in the hard X-ray regime, meaning that HPGe detectors exhibit substantially higher quantum efficiency for photon energies above 20~keV~\cite{MOHANTY2008186}, which allows an effective detection of weak signals in dilute samples containing heavy elements~\cite{Cole2001RareEarthEXAFS}. Additionally, Germanium’s smaller bandgap ($\sim$0.67~eV) enables finer energy resolution by generating more charge carriers per absorbed photon, improving the resultant signal to background/noise ratio (SNR)~\cite{FARROW1995567,dennis2019}.
The monolithic design avoids boundaries between separate crystals, which can cause signal variations due to gaps or misalignment. Using high-purity germanium reduces crystal defects and, hence, reduces charge-carrier trapping. As a result, the detector achieves high energy resolution and good signal response across the entire active area~\cite{Hexgonal_pad_N_tartoni}, thereby attaining qualities critical for spectroscopy applications.
Great care is required during fabrication, as such detectors require ultra-high purity and operate in a cryogenic environment (near 77~K) to suppress thermal noise. 

Building on this motivation,\textit{ Work Package 2} of the European LEAPS-INNOV~\cite{LEAPSINNOV2020} project has taken a step towards developing a new generation of HPGe detectors~\cite{F_ORSINI_FIRSTARTICLE, Goyal_2025, Gimenez_2025}. We have developed two such detectors with 10-element monolithic sensors, featuring pixel areas of 5~mm$^2$ and 20~mm$^2$. These detectors should offer the capability of handling high throughput in a broad X-ray energy range, while maintaining a reliable energy resolution aiming $\sim$180~eV FWHM at $\sim$5.9~keV, within the integration time window/Peaking Time (PT) of 1~$\mu$s~\cite{Goyal:2025oph}. In addition, the Digital Pulse Processing technique (DPP) will further improve detector performance by electronically rejecting charge-sharing events, which are responsible for degrading the resultant spectral resolution~\cite{QuantumXspress3} due to incomplete charge collection.
\par
This article presents the full development and characterization of the first HPGe prototype.
In Section~\ref{sec:DetGeom}, we describe the detector design, including the monolithic design of the sensor, the cryogenic cooling, mechanical architecture, and TETRA ASIC-based new front-end electronics.
Section~\ref{sec:det_charac} discusses the laboratory-based characterization with detailed analysis of signal rise time, Power spectral density, fluorescence measurements with metallic foil samples, and surface scanning using a micro spot beam. In Section~\ref{sec:beamtests}, we discuss the first on-beam evaluation of the detector, followed by a detailed analysis of defects observed in several pixels. We provide a phenomenological model to derive the depth of these defects.
Finally, we present a summary of detector performance to date and outline future directions for performance improvement.

\section{The XAFS-DET germanium detector}
\label{sec:DetGeom}
\begin{figure}[htb!]
\centering
\includegraphics[width=\textwidth]{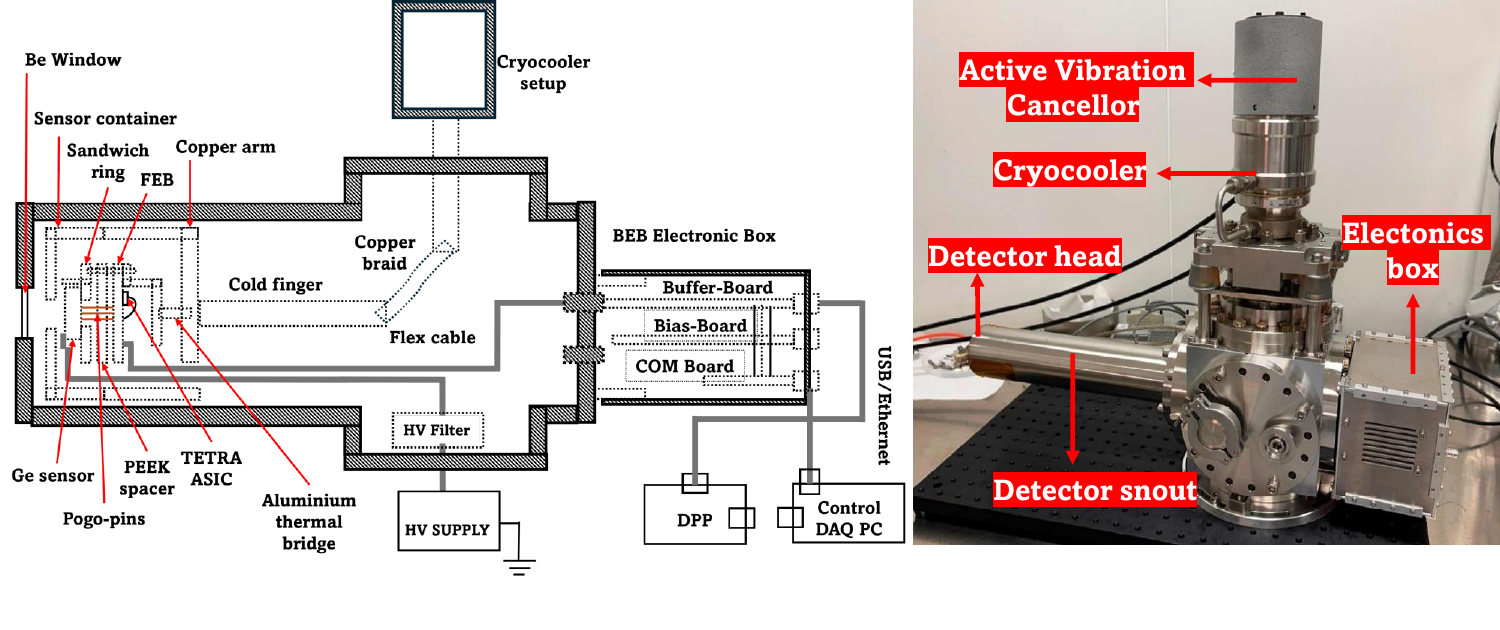}
\caption{Schema (left) and photo (right) of XAFS-DET detector, described in detail in the text.}
\label{Fig:Detector_Schema}
\end{figure}

The XAFS-DET detector, shown in Figure~\ref{Fig:Detector_Schema}, is composed of a compact detector head, a detector body, and ancillary elements. Other synchrotron spectroscopy detectors typically follow the same schematic structure. The detector head, which includes the germanium sensor, the front-end electronics, and their interconnections, is under vacuum at cold temperature (77~$K$). The detector body comprises the cooling system, the cryostat with its flanges, and the back-end electronics. Ancillary elements include the digital pulse processor, voltage power supplies, and equipment control and monitoring.

\subsection{Sensor Design}
\label{sec:sensor_design}
\begin{figure}[htb!]
\centering
\includegraphics[width=\textwidth]{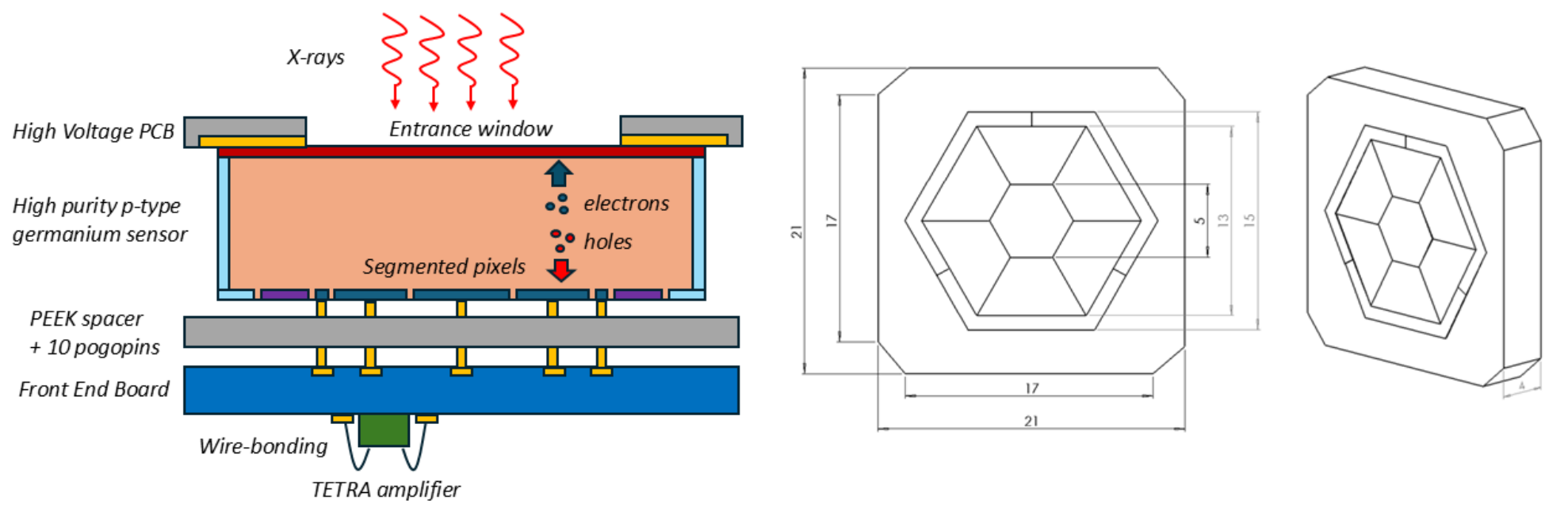}
\caption{Schematic representation of the germanium sensor (left) and sensor dimensions (right).}
\label{Fig:sensorschema}
\end{figure}

The sensor is a multi-element high-purity germanium sensor manufactured by Mirion Technologies (CANBERRA). A scheme of the sensor is shown in Figure~\ref{Fig:sensorschema}. It has a square shape (20~mm length) and a thickness of 4.1~mm. The sensor is a p-type germanium bulk operating in hole-collection mode, which is expected to reduce charge trapping relative to electron collection. The front side (facing the incident X-rays) is phosphor-implanted (n-type doped).
Whereas, the back side is boron-implanted (p-type doped) and is segmented into a central hexagonal pixel, six trapezoidal pixels situated in a ring around the central one, and three outer ring-shaped pixels, which serve as ancillary pixels to reject charge-sharing events by sharing the incident charge with the guard ring. The dimensions of hexagonal and trapezoidal pixels are defined to achieve an area of 20 mm$^2$ per pixel. Aluminium contacts were evaporated on both sides, and the inter-pixel spacing provides a simulated insulation value of 10~T$\Omega$. A positive potential difference is established by applying a high voltage to the frontside window contact via a high-voltage Rogers PCB, while backside pixels are connected to front-end board pads via copper-gold-plated pogo pin connectors (Model P706 from Peak Test, 6.9~mm long with 0.2~mm-diameter tips). The full depletion voltage is around +33 V, and we operated the detector in the range of  +75 to +300 V during characterization. We measured the leakage current of each pixel to be less than 1~pA at +100V while operating at liquid nitrogen temperature.

The germanium sensor is caged in a sensor head assembly. It consists of an aluminum container (26~mm radius) with a square hole (14~mm length), covered by an infrared window. The sensor is situated between the high-voltage PCB, a PEEK spacer that keeps the pogo pins in place on the backside, and the front-end board, whose preamplifiers and wire bonds are protected by an aluminum cover. Stainless steel screws keep all pieces together.  A 1~mm thick tungsten collimator is screwed to the container's front side.
\subsection{Electronics}
\label{sec:Electronics_FE_BE}

The readout electronics is composed of three parts, all designed and manufactured by XGLab-BRUKER: the TETRA ASIC, the front-end board (FEB), and the back-end board (BEB).

\begin{figure}[htb!]
\centering
\includegraphics[width=\linewidth]{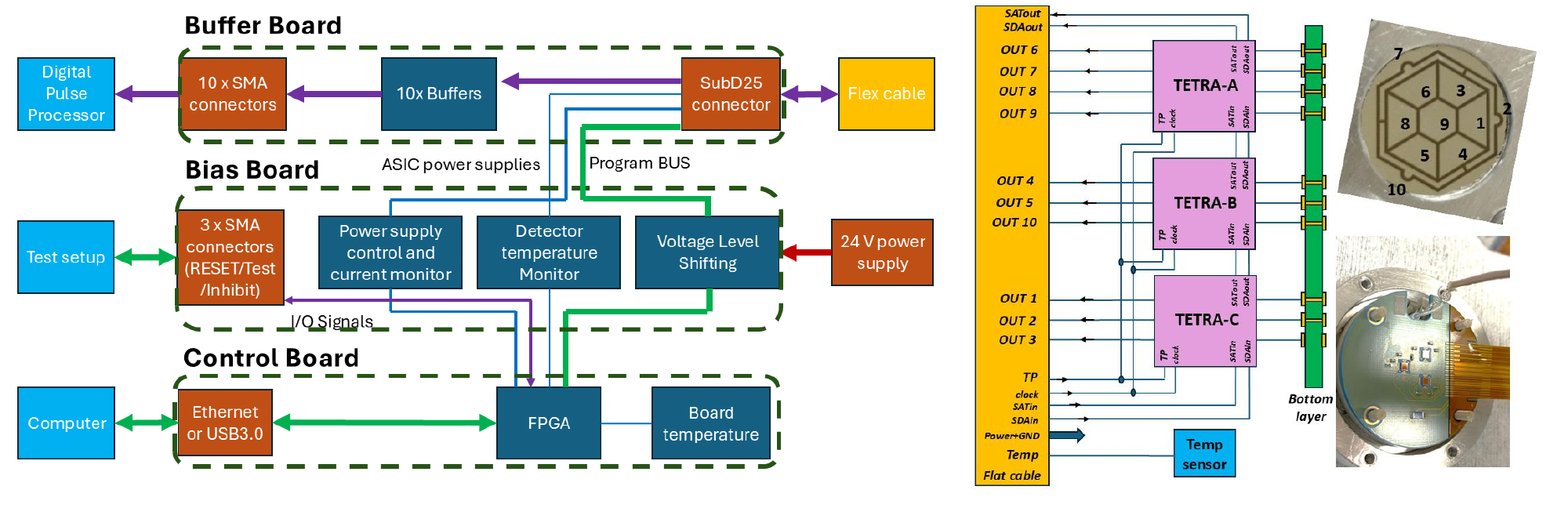}
\caption{Conceptual schema of the three BEB and the FEB, described in detail in the text.}
\label{Fig:bebschema}
\end{figure}

TETRA ASIC is a four-channel charge-sensitive (CSA) preamplifier working in hole collection mode with synchronous reset~\cite{6154396}. The preamplifier gain setting (low, medium, or high), which varies for feedback capacitances of 29, 58, and 174~fF, respectively, Reset features (threshold, width, and external trigger), and channel disabling can be configured by the user via a 32-bit shift register. All four channels share a test-pulse input to characterize signal fall time and crosstalk between channels without the sensor, as well as to simulate input leakage current~\cite{CMOS_CSA}.
The FEB is a circular board (20~mm radius, 1.4~mm thick) made of ceramics, as shown in Figure~\ref{Fig:bebschema} (left). It links the ten pogo pin pads (0.4~mm radius), on its front side, to ten input channels of three TETRA ASICs (two with three channels enabled, one with all four channels enabled) located on its back side. The FEB back side also houses a thermistor. Output signals and TETRA power supplies are routed to the BEB via a 60~cm-long flex kapton PCB cable with a Sub-D 25 connector. The cable is glued to the FEB back surface, and electrical continuity is achieved via wire bonding, while the Sub-D 25 connector is screwed into the back-flange feedthrough of the vacuum cube.

The BEB comprises three PCB boards (buffer, bias, and control) designed to bias and configure three TETRA ASICs and buffer 10 analog channels. A conceptual schema is shown in Figure~\ref{Fig:bebschema} (right). The buffer board is directly connected to the subd-25 feedthrough to minimize the path of ASIC outputs. The ten signals are buffered and routed to ten SMA connectors, which are connected to the Digital Pulse Processor via coaxial RF cables. The bias board supplies power to itself and the other two boards, applies voltage-level shifting to the ASIC program bus, and acquires the FEB thermistor signal. The bias board also includes three SMA connectors: two inputs for the external reset pulse and the test pulse, and one output for the inhibit reset signal. The control board monitors and controls BEB features via an FPGA unit and communicates with the user via USB 3.0 or Ethernet.

\begin{figure}[ht]
    \centering
    \begin{subfigure}[b]{0.48\textwidth}
        \includegraphics[width=\textwidth]{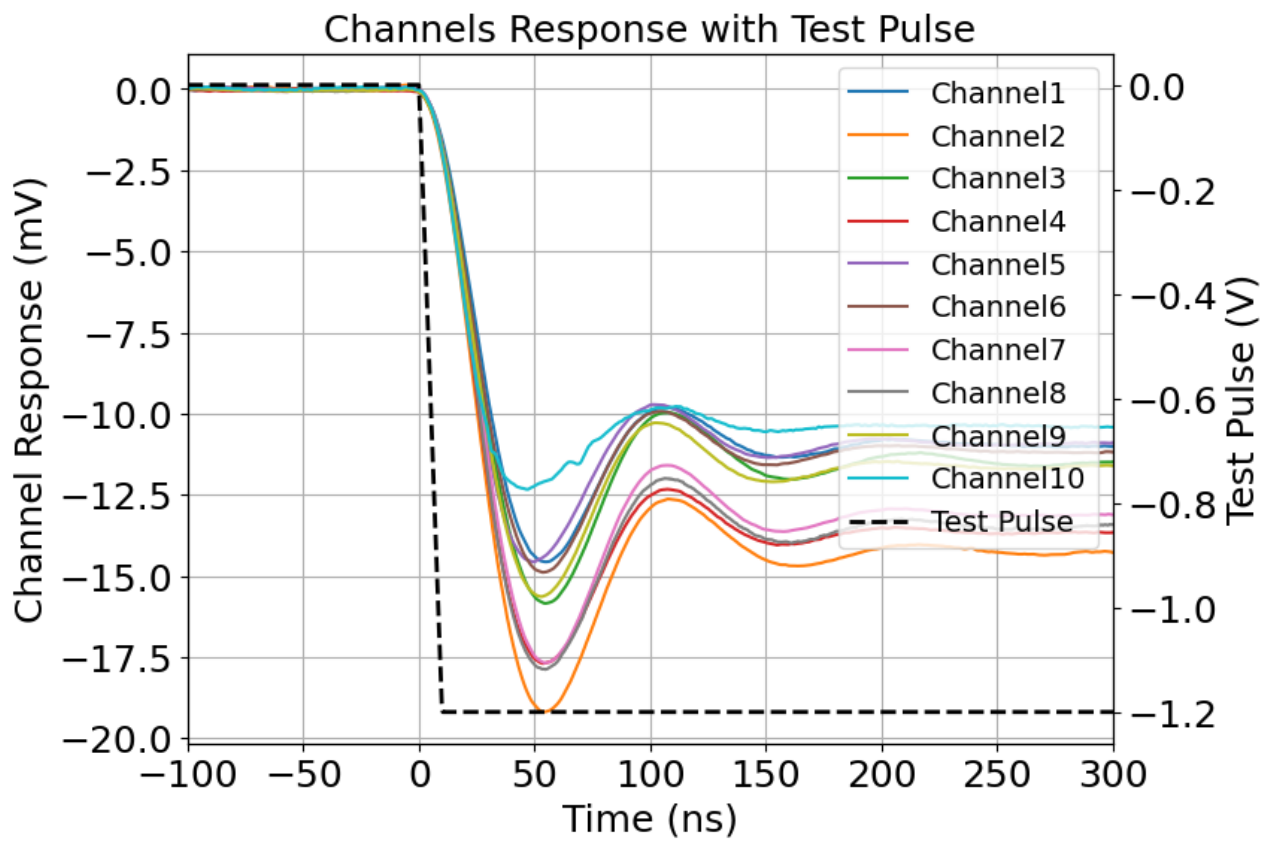}
        \caption{}
        \label{Fig:highgain_allchannels_falltime}
    \end{subfigure}
    \hfill
    \begin{subfigure}[b]{0.48\textwidth}
        \includegraphics[width=\textwidth]{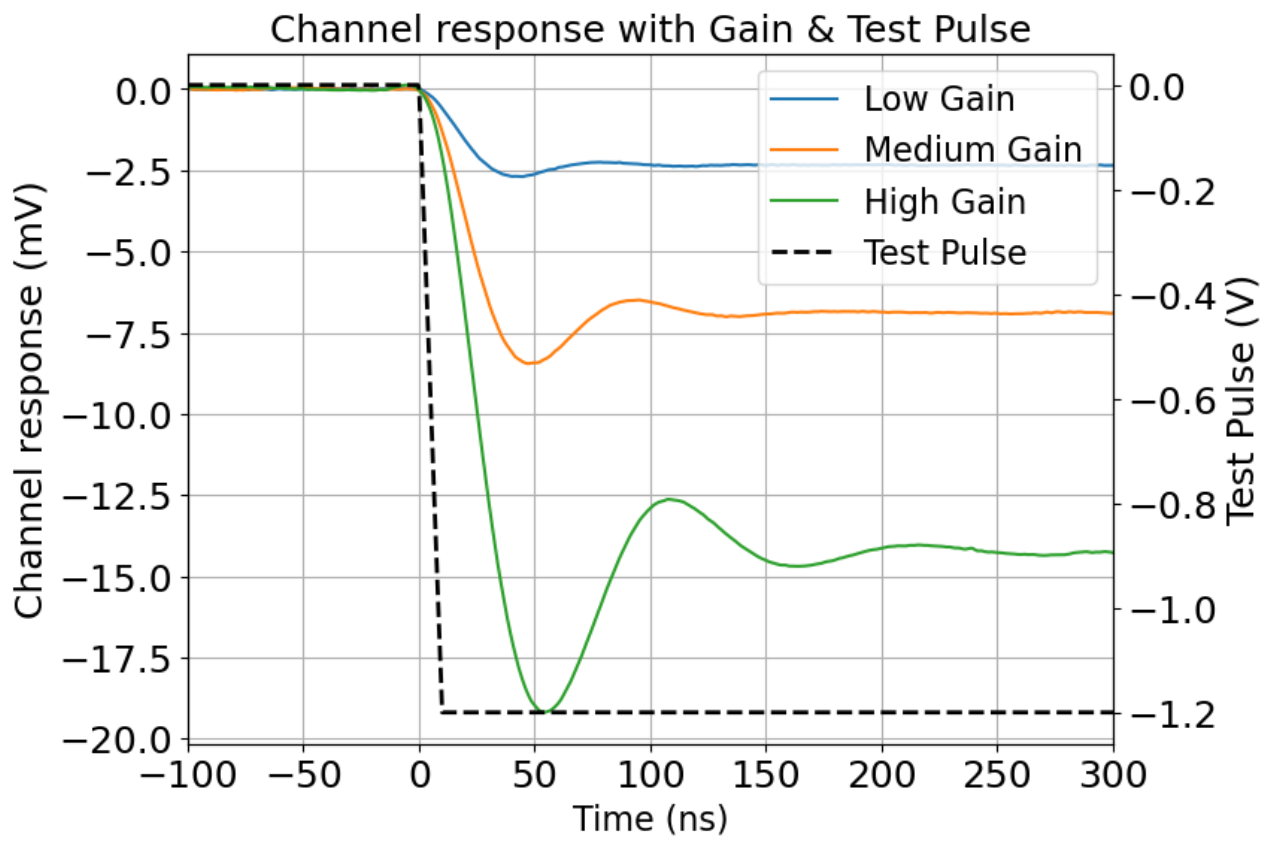}
        \caption{}
        \label{Fig:allgain_1channels_falltime}
    \end{subfigure}

\caption{(a) Test pulse response for the 10 channels of the FEB for high gain (b) and for channel 2 for high, medium and low gain settings.}
\end{figure}

The electronics performance is summarized in Table~\ref{tab:electronics_performance}. The step function response without the sensor was characterized at room temperature (23-25°C) in terms of signal gain and fall-time (10\% to 90\% of the amplitude) for the three gain settings and for all channels. These features were measured from the mean waveform generated by a squared-test pulse (1~V amplitude, 10~ns rising edge), as illustrated in Figure~\ref{Fig:highgain_allchannels_falltime}. Fall-time values of 19-23~ns were measured for all gains, and we verified that the gain ratios high-medium and medium-low were 2.0 and 3.0, see Figure~\ref{Fig:allgain_1channels_falltime}. We also quantified absolute gain values during subsequent laboratory tests using the sensor. In a second step, the range of values for reset duration (0.7-5.9~$\mu$s) and threshold (-2.3-1.3~V) was measured. The reset period was longer than 4-5~sec, for a negligible leakage current. The period was shorter (around 23, 41, and 116 ms for high, medium, and low gain settings, respectively) with the leakage emulator enabled. Finally, the Equivalent noise charge (ENC) for peaking times between 0.1 and 10~$\mu$s was measured, yielding values of 36-45 and 29-35 electrons for 1.0 and 10~$\mu$s, respectively, within specifications. This series of tests was repeated at cold temperature (113~K), as discussed in~\cite{Goyal:2025oph}.

\begin{table}[htb!]
\centering
\caption{Summary of key electronic performance parameters from the Site Acceptance Tests.}
\label{tab:electronics_performance}
\begin{tabularx}{\textwidth}{lXX}
\toprule
\textbf{Parameter} & \textbf{Target Specification} & \textbf{Measured Result} \\
\midrule
\multicolumn{3}{c}{\textbf{TETRA ASIC Performance}} \\
\midrule
Channel Gain (mV/keV) & 0.36, 1.1, 2.2 & -- \\
Risetime (at 20°C) & < 22~ns & 19 - 23~ns \\
Reset Period & > 1~sec & 4-5~sec \\
Reset Duration & < 2.0~$\mu$s & 0.7 -- 5.9~$\mu$s\\
\midrule
\multicolumn{3}{c}{\textbf{Full Chain Performance (FEB + BEB)}} \\
\midrule
ENC ( 1 $\mu$s PT )& < 45 electrons & 36-45 electrons \\
ENC ( 10 $\mu$s PT ) & < 35 electrons & 29-35 electrons \\
Buffer Board Crosstalk & < 0.2\% & ~0.3\% \\
Vacuum Outgassing Rate (FEB C2) & < $10^{-7}$ mbar $\times$ l/s & $7 \times 10^{-7}$ mbar $\times$ l/s \\
\bottomrule
\end{tabularx}
\end{table}

\subsection{Cooling and mechanical design}
\label{sec:Cooling_mechanical}
The detector cooling system is based on an electrically driven Stirling-cycle cryocooler (model CryoTel CT, developed by Sunpower). This closed system works by compressing and expanding a fixed volume of Helium gas, enabling efficient, continuous cooling without refilling. Although cryocooler may induce high mechanical vibrations to the cold finger, these vibrations have been compensated by mechanical damping and thermal decoupling elements, inspired by the Maroon-X detector system~\cite{Seifahrt:2018as}.
\begin{figure}[htb!]
\centering
\includegraphics[width=0.65\linewidth]{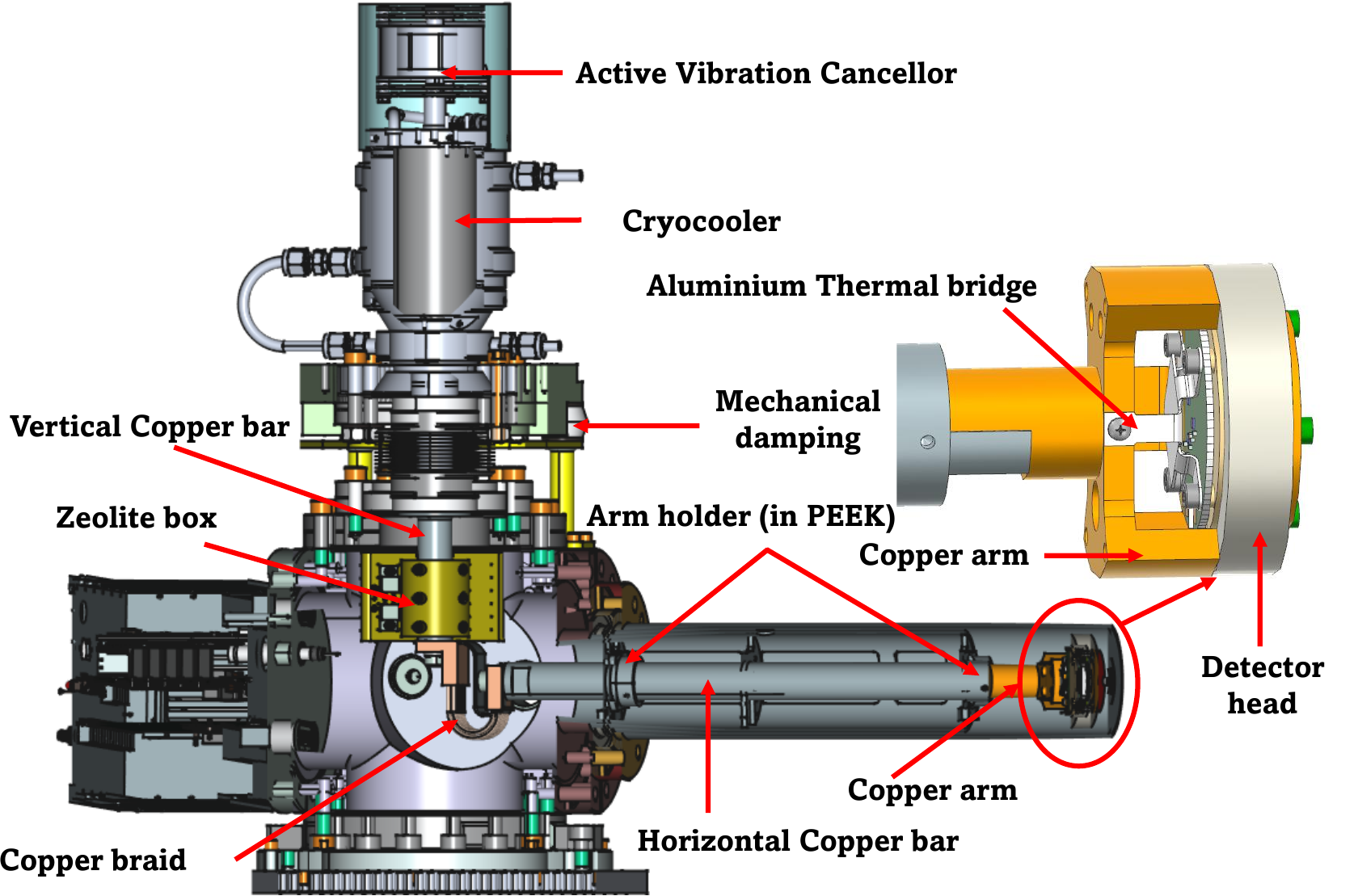}
\caption{Mechanical design of cooling chain and the associated components.}
\label{Fig:cooling_mechanicaldesign}
\end{figure}
The aluminum container, a copper arm holder, a horizontal copper cooling pipe, a thermal strap, and a vertical copper braid to the cryocooler evacuate the dissipated power from the front-end board~\cite{Quispe:2024wis}.The pads of Indium foils link each interface. The cooling pipe is supported by two PEEK arm holders inside the inner cylinder, with small supports to minimize thermal losses. Two thermometers, glued to the vertical copper braid and the copper arm, monitor the temperature. The second one works as a feedback device for the cryocooler system. A zeolite box is attached to the vertical copper braid piece, which maintains vacuum integrity in static vacuum by adsorbing residual gases using pre-activated zeolite treated under vacuum at high temperatures. The box is equipped with a heating element and a thermometer, which are used to regenerate zeolites once they are saturated. The cooling chain of the detector is showcased using Figure\ref{Fig:cooling_mechanicaldesign}
.

The cryocooler is mounted on a vacuum cube that holds the detector head in a front-facing tube. The cube's three CF100 flanges provide dedicated feedthroughs for electronics (Sub-D 25), cooling thermometers (SubD9), and HV bias. A separate flange houses the pumping port, vacuum gauge, and a SubD9 for the zeolite heater. A low-pass RC filter with a 6.5~Hz cut-off frequency conditions the HV signal to the sensor. A turbo pumping system achieves a vacuum of $2 \times 10^{-7}$~mbar, with residual gas analysis (\textit{MKS model}) confirming only trace amounts of air and water.\\
 Induced vibrations at the sensor head were measured using an Active Vibration Cancellation (AVC) system (\textit{model VSM101}) produced by ifm Electronics. Acceleration values of 0.04-0.07~m/s$^2$ at 1-2 frequencies above 50~Hz were measured in the three axes, which are within specifications.

\section{Detector characterization in the laboratory}
\label{sec:det_charac}

The laboratory characterization of the XAFS-DET detector included:  
(i) measurement of the signal rise time and preamplifier gain,  
(ii) evaluation of the power spectral density,  
(iii) testing with various fluorescence samples using an X-ray generator, and  
(iv) micro-beam surface scans.  \\
The procedures and results for each test are discussed in detailed in the following subsection.
\subsection{Mean Signal risetime and preamplifier gain}
\label{sec:det_charac_risetime}

The signal rise time and preamplifier gain were characterized using a \textsuperscript{55}Fe radioactive source emitting 5.9~keV $(K_{\alpha1})$ and 6.4~keV $(K_{\beta1})$ X-rays. We operate the detector under low-flux conditions (approximately 10~kcps), and subsequently we acquire 10$^{5}$ individual signal waveforms using a \textit{Lecroy} digital oscilloscope operating at 0.5~Gs/s.  
We performed the measurements with bias voltages varying from +50~V to +330~V in 10~V increments.

\begin{figure}[ht]
    \centering
    \includegraphics[width=0.6\textwidth]{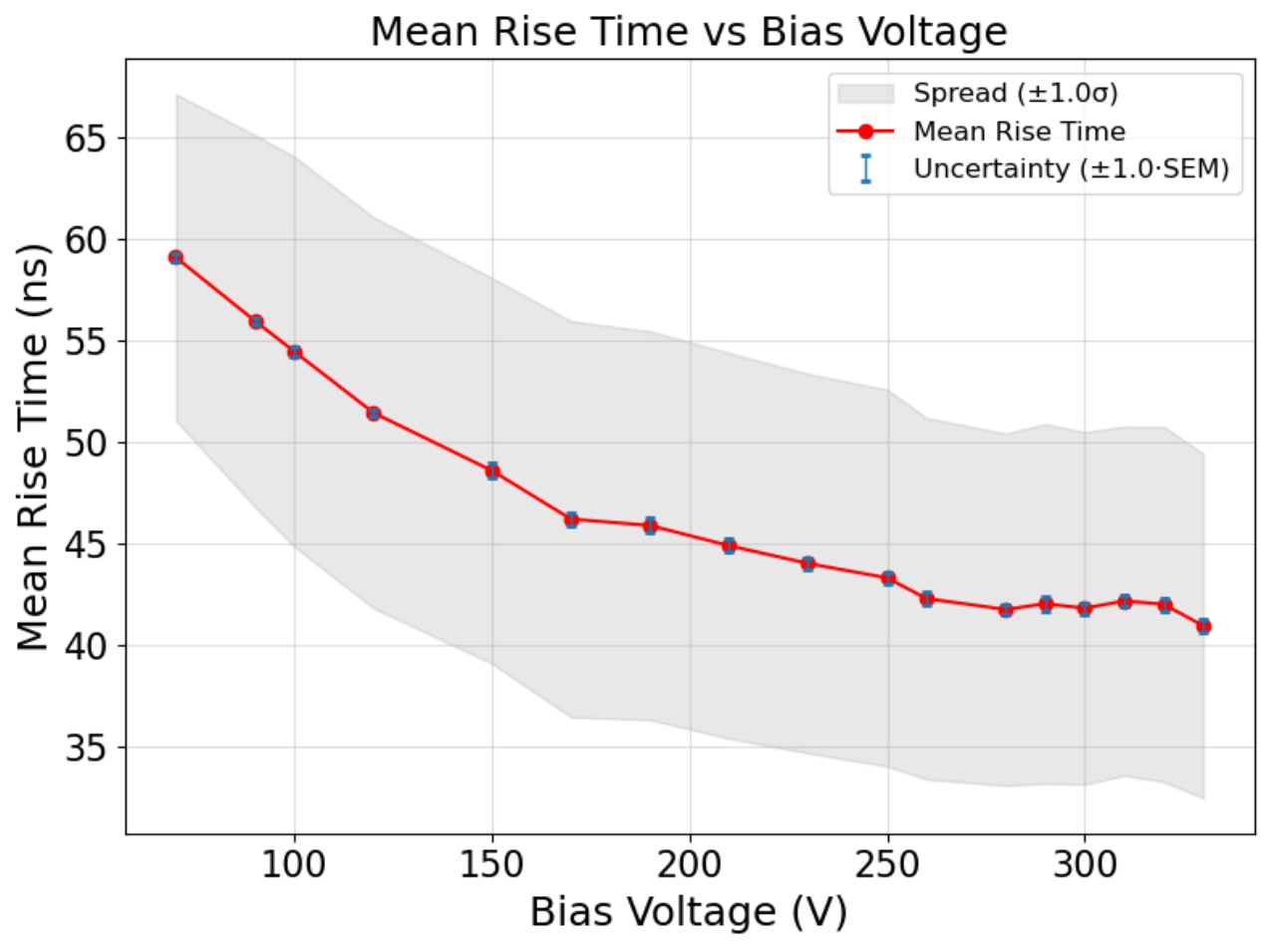}
    \caption{Dependence of the mean and standard deviation (error bars) of the risetime distribution with Detector bias voltage.}
    \label{fig:risetimevsbiasvoltage}
\end{figure}

\begin{figure}[ht]
    \centering

    \begin{subfigure}[b]{0.48\textwidth}
        \centering
        \includegraphics[width=\textwidth]{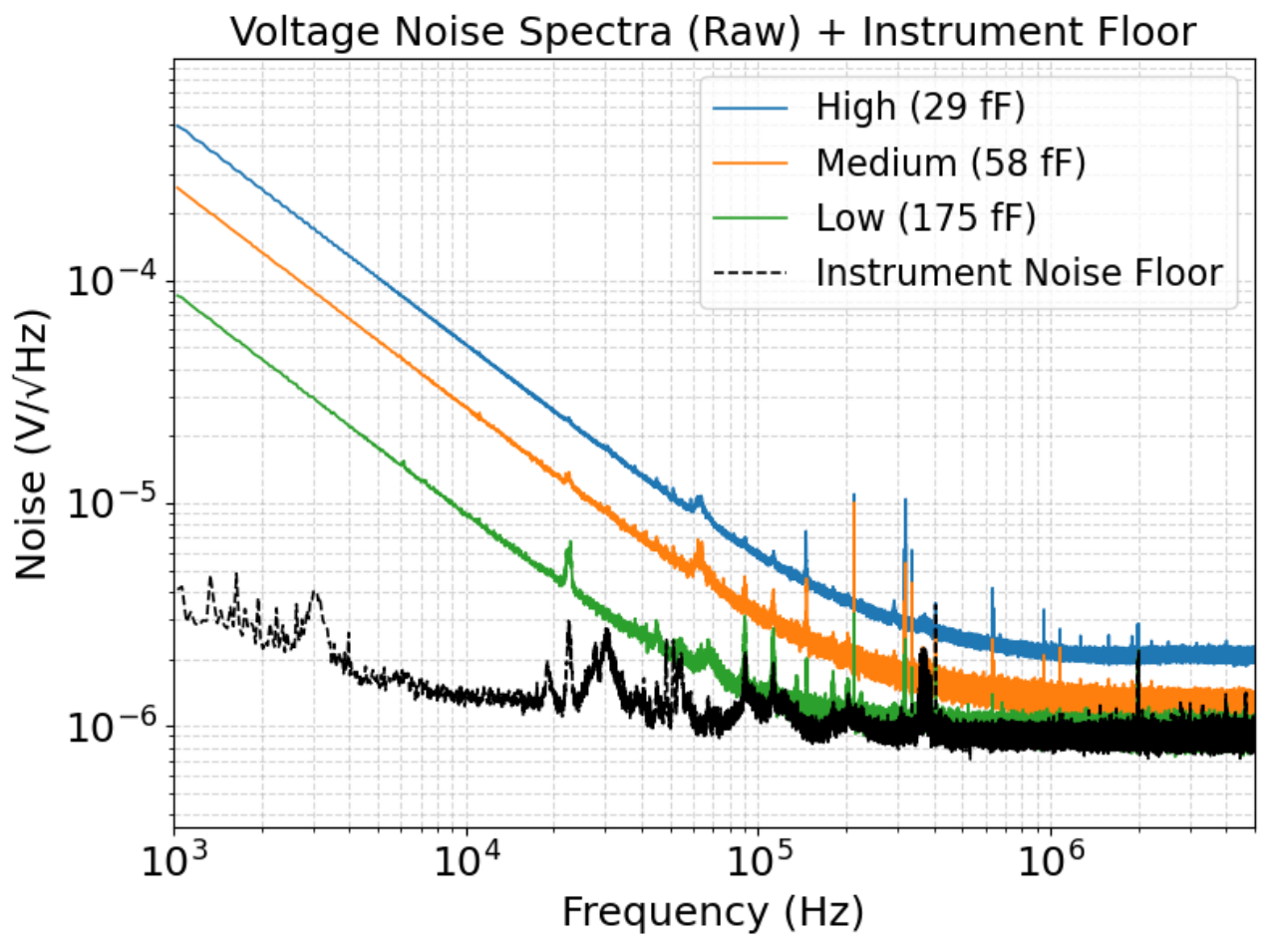}
        \caption{.}
        \label{fig:psd_tetra}
    \end{subfigure}
    \hfill
    \begin{subfigure}[b]{0.48\textwidth}
        \centering
        \includegraphics[width=\textwidth]{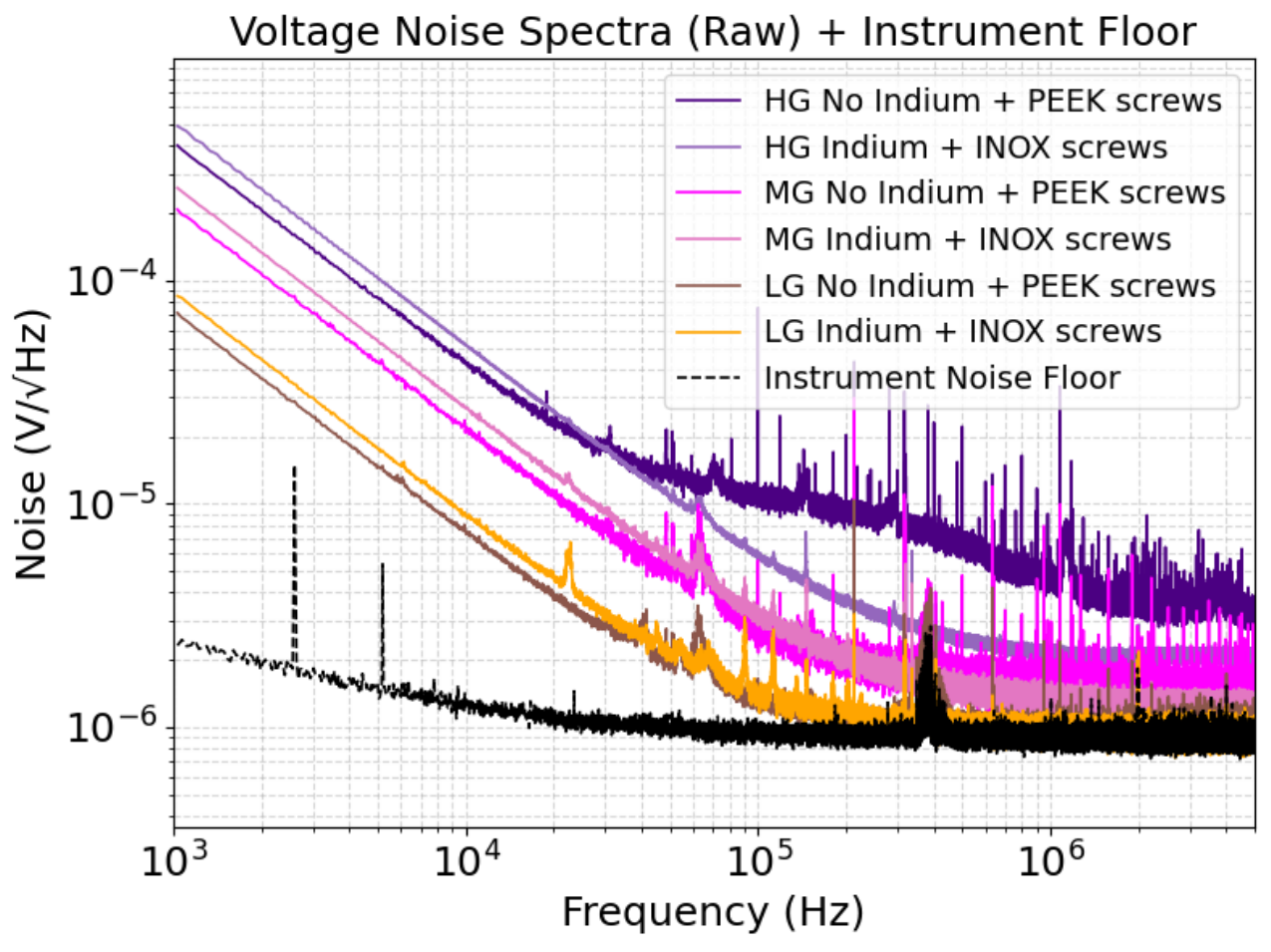}
        \caption{}
        \label{fig:psd_ind_peek_inox}
    \end{subfigure}

    \caption{(a) Power Spectral Density (PSD) for the central pixel and the three TETRA gain values (b) PSD analysis with two different configurations of with and without the Indium foils on the sensor surface, and front-end interconnection via PEEK/INOX screws.}
    \label{fig:psd_configs}
\end{figure}

Each recorded waveform was processed to extract the signal amplitude and rise time. The distribution of these parameters is analyzed as a function of bias voltage. The mean pulse amplitude was used to determine the preamplifier gain, yielding values between 1.99 and 2.06~mV/keV with a systematic uncertainty of $\pm$0.02~mV/keV, which are within design specifications.  

As shown in Figure~\ref{fig:risetimevsbiasvoltage}, the rise time distributions shift progressively toward shorter values as the bias voltage increases. This trend reflects the expected enhancement in hole drift velocity with stronger electric fields, reaching a minimum rise time of about 42~ns in the 280–330~V range.

\subsection{Power Spectral Density Measurements}
\label{sec:noise}
    The Power Spectral Density (PSD) measurements are done to evaluate  how the noise power of a signal is distributed across different frequencies, and can be obtained by taking the Fourier transform of the measured voltage signal 
\(v(t)\), over the acquisition time 
\(T\).
The PSD can be expressed as:
    \[
    \text{PSD}(f) = \frac{1}{T} \left| \mathcal{F}[v(t)] \right|^2
    \]
    The corresponding voltage noise spectral density can be  expressed as:
    \[
    S_V(f) = \sqrt{\text{PSD}(f)} \quad [\text{V}/\sqrt{\text{Hz}}]
    \]
    We measured the PSD over the frequency range of 100~Hz to 50~MHz using a fast digitizer (\textit{Model DT9862}, by Data Translation, triggered 100~ms after the RESET signal to capture noise during the signal-processing window. Measurements were performed on the central pixel across all available TETRA ASIC gain settings. As shown in Figure~\ref{fig:psd_tetra}, the noise level above 100~kHz is approximately $10^{-6}$~V/$\sqrt{\text{Hz}}$, limited by the digitizer sensitivity. Several narrow 1/$f$ components appear between 140 and 320~kHz. At lower frequencies, the PSD rises due to voltage noise contributions shaped by the detector input capacitance (estimated at $\sim$4~pF, including the pogo-pin and front-end board input pads).\par
    Beyond simply characterizing the ASIC gain behavior, this analysis also enabled evaluation of different mechanical and electrical configurations of the system. We compared measurements with the back-end electronics housing open and closed, to evaluate how shielding the  DC–DC converter present in the bias board is influencing the noise coupling. In addition, we also compared the configurations combining the use of indium foil pads and two different screws material (PEEK vs. stainless steel) used for sensor and front-end PCB interconnection. As shown in Figure~\ref{fig:psd_ind_peek_inox}, the configuration with indium foil and INOX (Stainless steel) screws consistently resulted in lower noise levels across three gain configurations.
\subsection{Fluorescence Measurements with lab X-ray generator source}
\label{sec:labxraygentests}
To characterize the detector response as a function of photon flux, fluorescence spectra were acquired with samples of $^{55}$Mn, $^{63}$Cu, $^{119}$Sn, and $^{137}$Ba using a laboratory X-ray generator. The generator was equipped with a molybdenum (Mo) anode, operated at a maximum voltage $\&$ current of 50~kV and 40~mA, respectively. The maximum photon energy generated is determined by the applied voltage, reaching up to 50~keV in this setup.
A set of multiple spectra was recorded across a range of peaking times (PT) from 0.5 to 12~$\mu$s, with acquisition windows of 20–30~s per spectrum. The input count rate (ICR) was systematically varied from 6~kcps to 1.2~Mcps.
\subsubsection{Energy Resolution $\&$ Count Rate Effects in Fluorescence Mode}
\label{sec:Energyresolution_ICR}
\begin{figure}[ht]
    \centering
    \begin{subfigure}[b]{0.48\textwidth}
        \includegraphics[width=\textwidth]{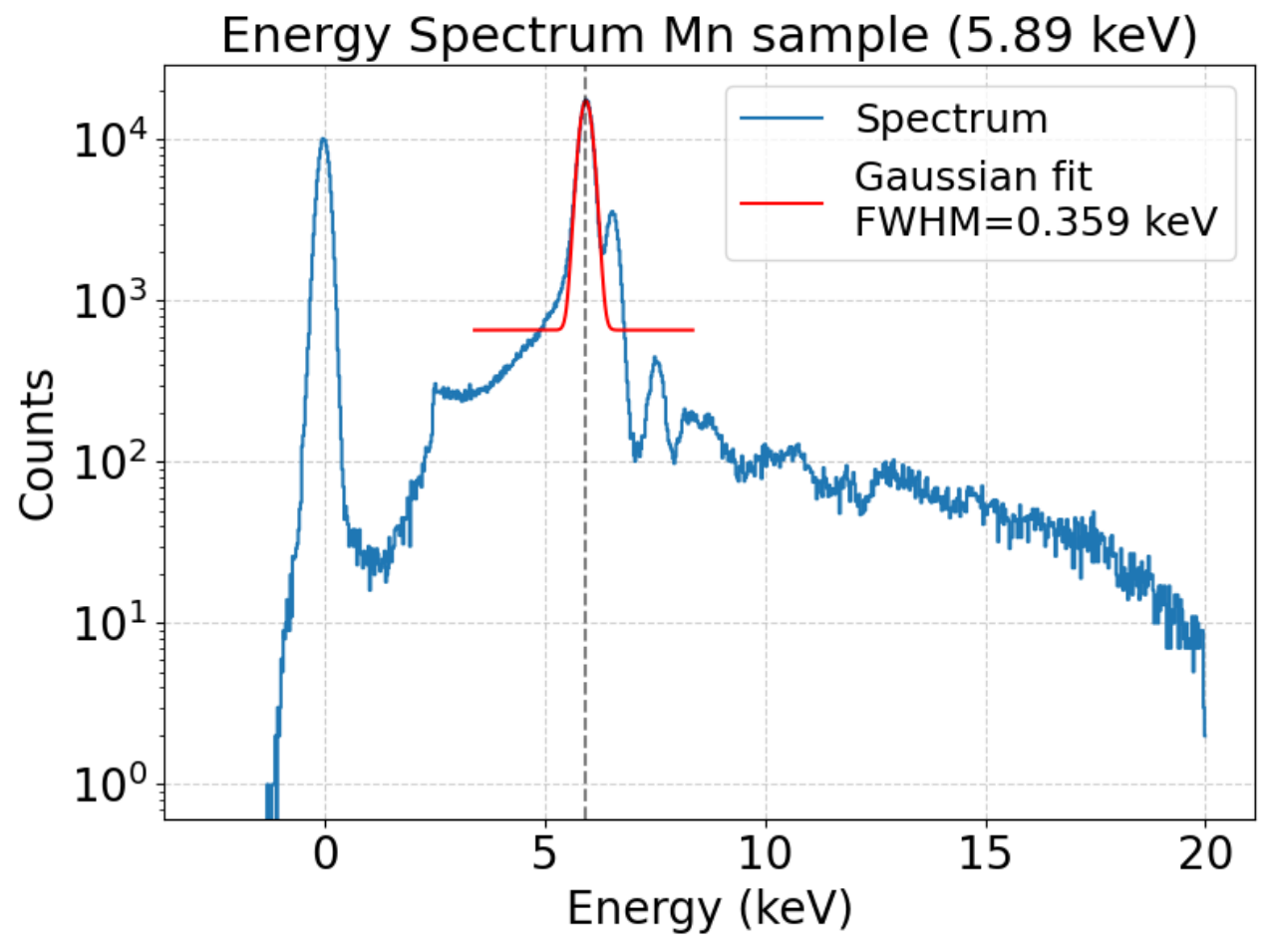}
        \caption{}
        \label{fig:mn_spec}
    \end{subfigure}
    \hfill
    \begin{subfigure}[b]{0.48\textwidth}
        \includegraphics[width=\textwidth]{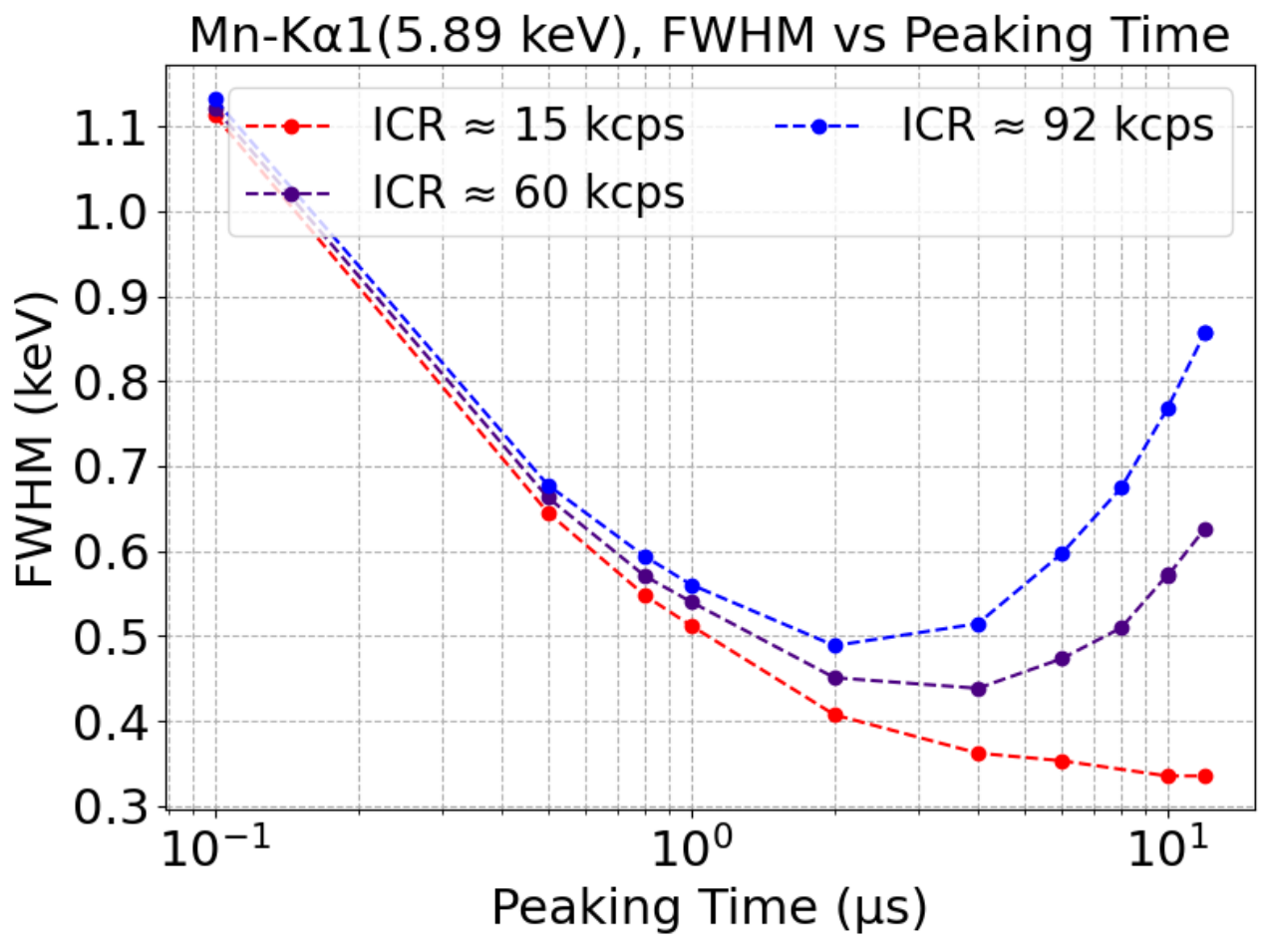}
        \caption{}
        \label{fig:mn_fwhm}
    \end{subfigure}

    \begin{subfigure}[b]{0.48\textwidth}
        \includegraphics[width=\textwidth]{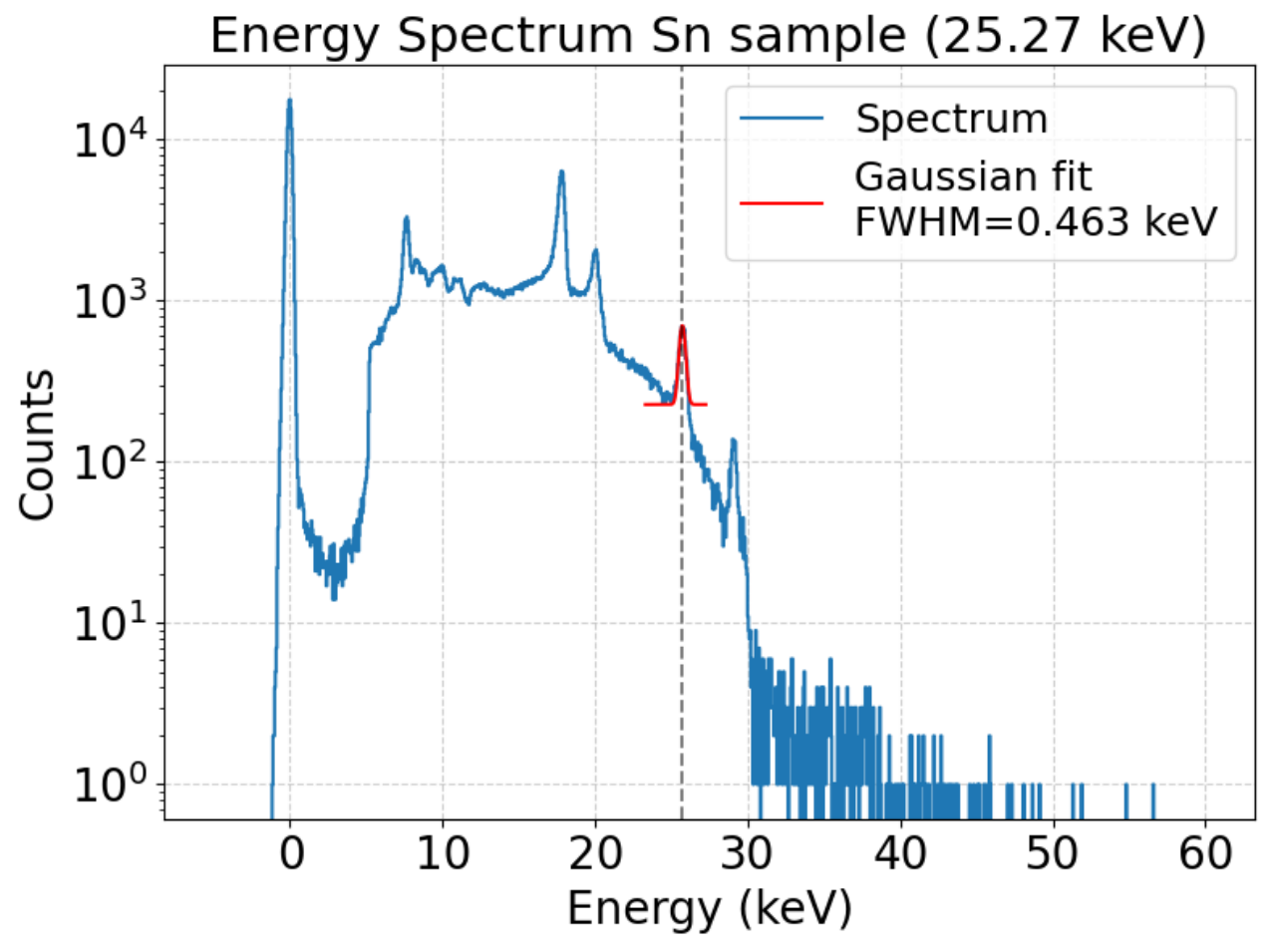}
        \caption{}
        \label{fig:sn_spec}
    \end{subfigure}
    \hfill
    \begin{subfigure}[b]{0.48\textwidth}
        \includegraphics[width=\textwidth]{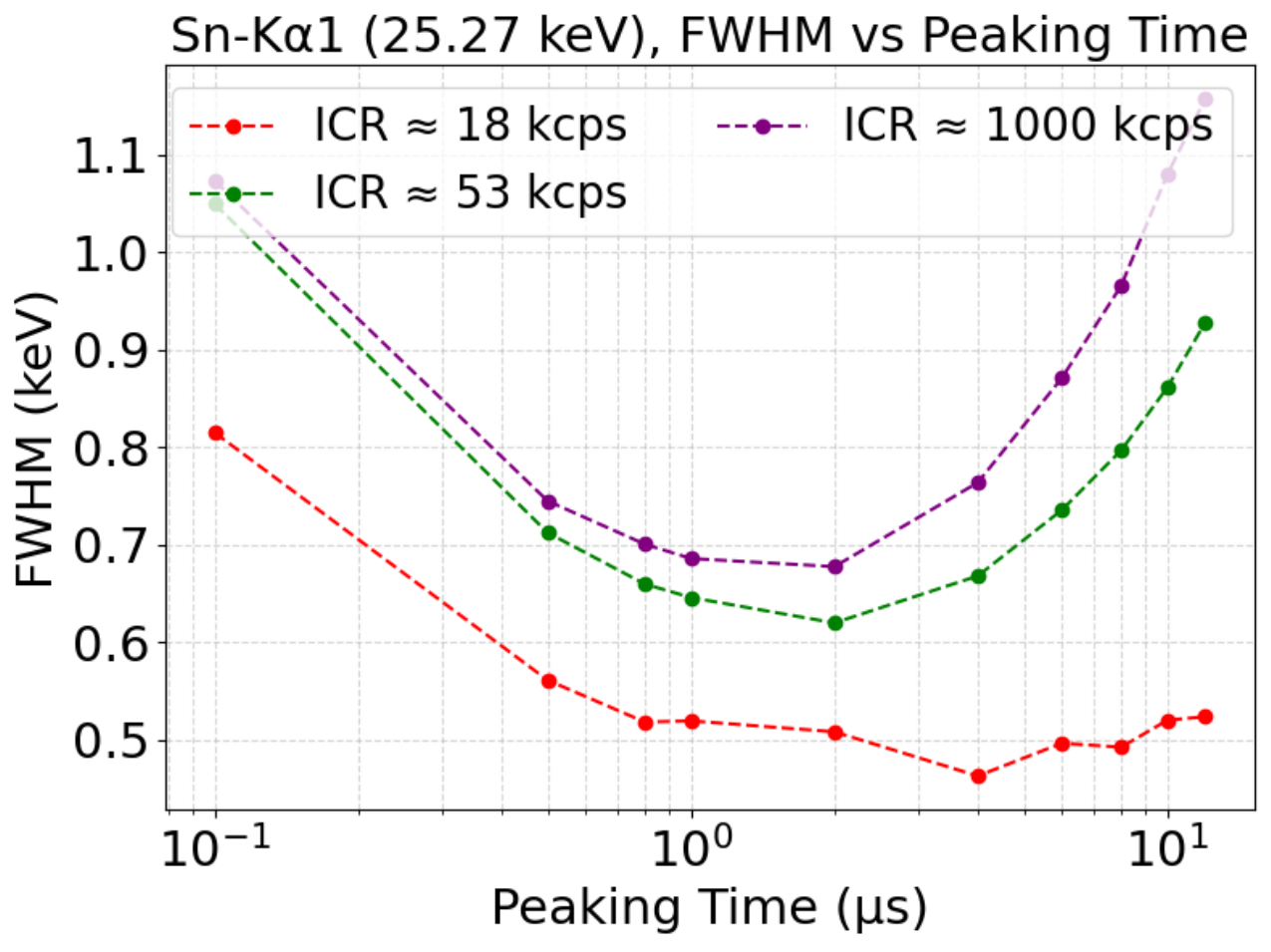}
        \caption{}
        \label{fig:sn_fwhm}
    \end{subfigure}

    \begin{subfigure}[b]{0.48\textwidth}
        \includegraphics[width=\textwidth]{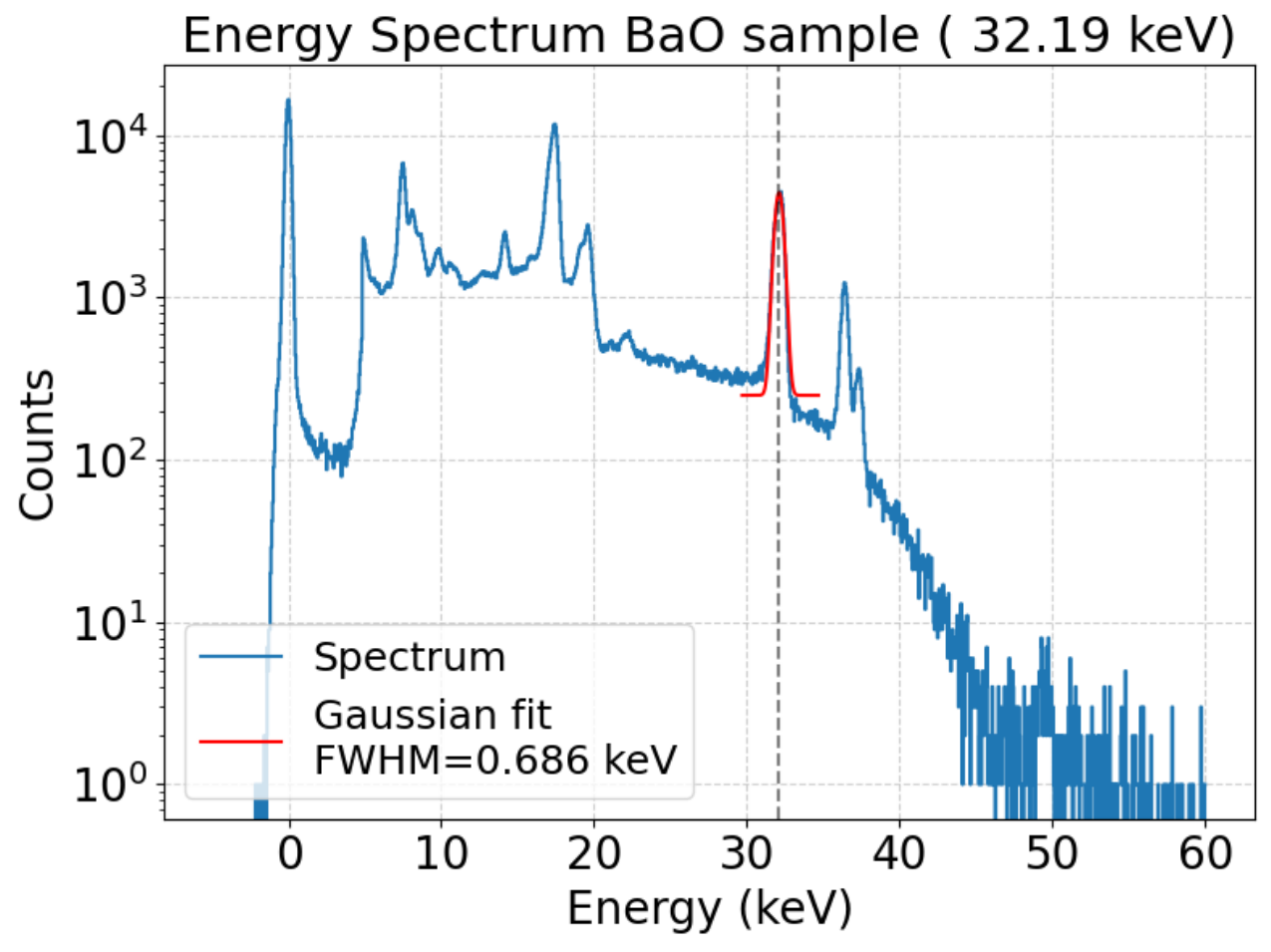}
        \caption{}
        \label{fig:ba_spec}
    \end{subfigure}
    \hfill
    \begin{subfigure}[b]{0.48\textwidth}
        \includegraphics[width=\textwidth]{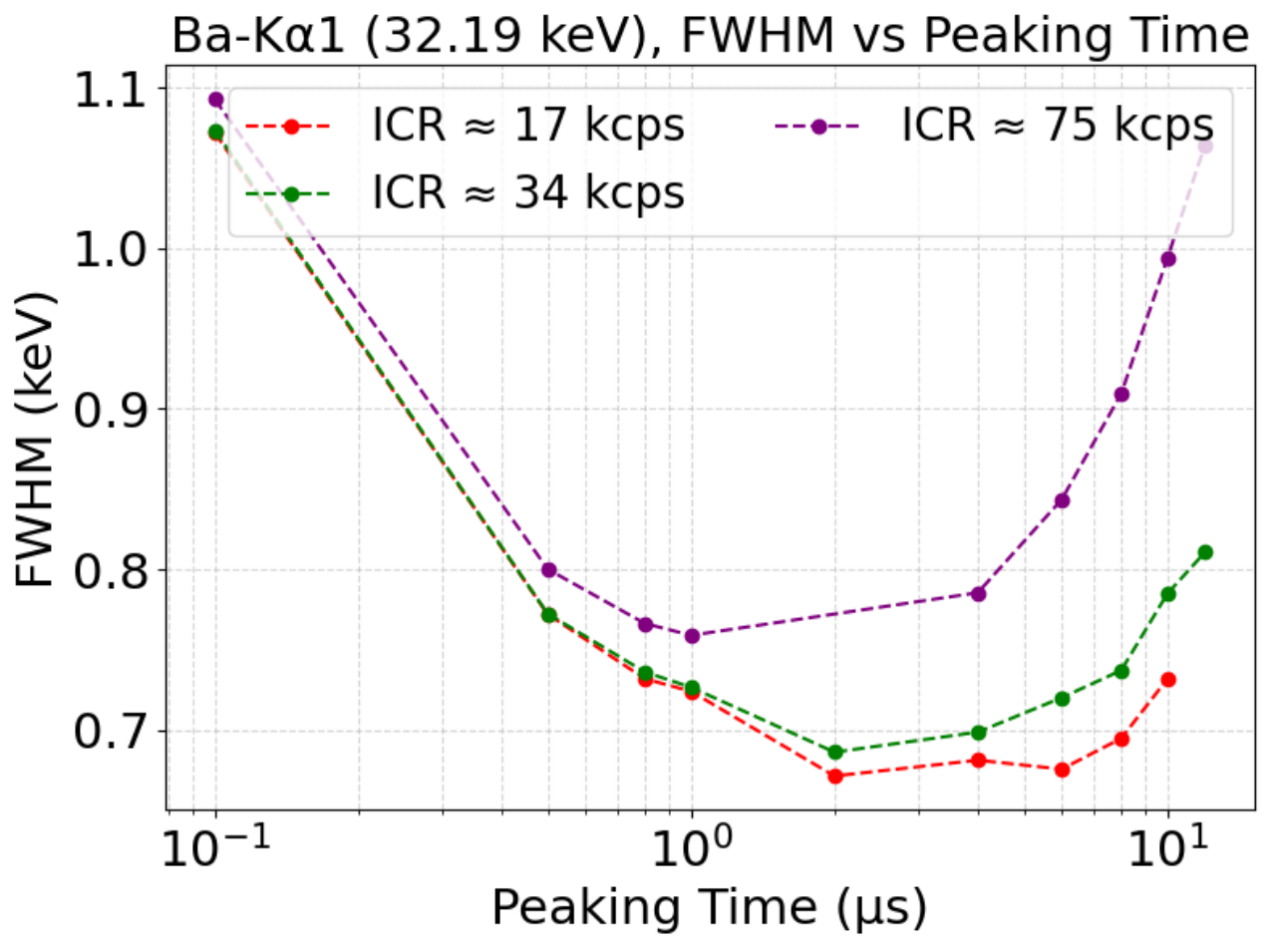}
        \caption{}
        \label{fig:ba_fwhm}
    \end{subfigure}

    \caption{
        Energy spectra and resolution performance for different fluorescence target foils.
        Left column: X-ray spectra from Mn (a), Sn (c), and BaO (e) targets. 
        Right column: Corresponding FWHM vs peaking time plots (b, d, f), under varying shaping times and count rates.
    }
    \label{Fig:fluorescence_PTvsFWHM}
\end{figure}

The sample foils used for characterization were selected to test and cover the three ASIC gain settings. The fluorescence spectra were calibrated using the characteristic X-ray lines at 5.9~keV(Mn-$K_{\alpha1}$), 25.27~keV (Sn-$K_{\alpha1}$), and 32.1~keV (Ba-$K_{\alpha1}$). The corresponding energy resolution are also showcased using Figures~\ref{fig:mn_spec},~\ref{fig:sn_spec},~\ref{fig:ba_spec}.
At low input count rates ($\sim$10~kcps), the detector achieved optimal resolution at shaping times between 4 and 6~$\mu$s, where the noise contribution was minimal, and ballistic deficit was minimal. As the ICR increased to moderate levels ($\sim$200–300~kcps), the optimal shaping time shifted slightly toward 2–4~$\mu$s, balancing resolution and event throughput. At high ICRs ($\sim$1.2~Mcps), resolution degraded significantly for PT values above 6~$\mu$s, while shorter PTs provided better resolution in a higher flux environment, as shown in Figures~\ref{fig:mn_fwhm},~\ref{fig:sn_fwhm},~\ref{fig:ba_fwhm}.
\begin{figure}[ht]
    \centering
    \begin{subfigure}[b]{0.49\textwidth}
        \includegraphics[width=\textwidth]{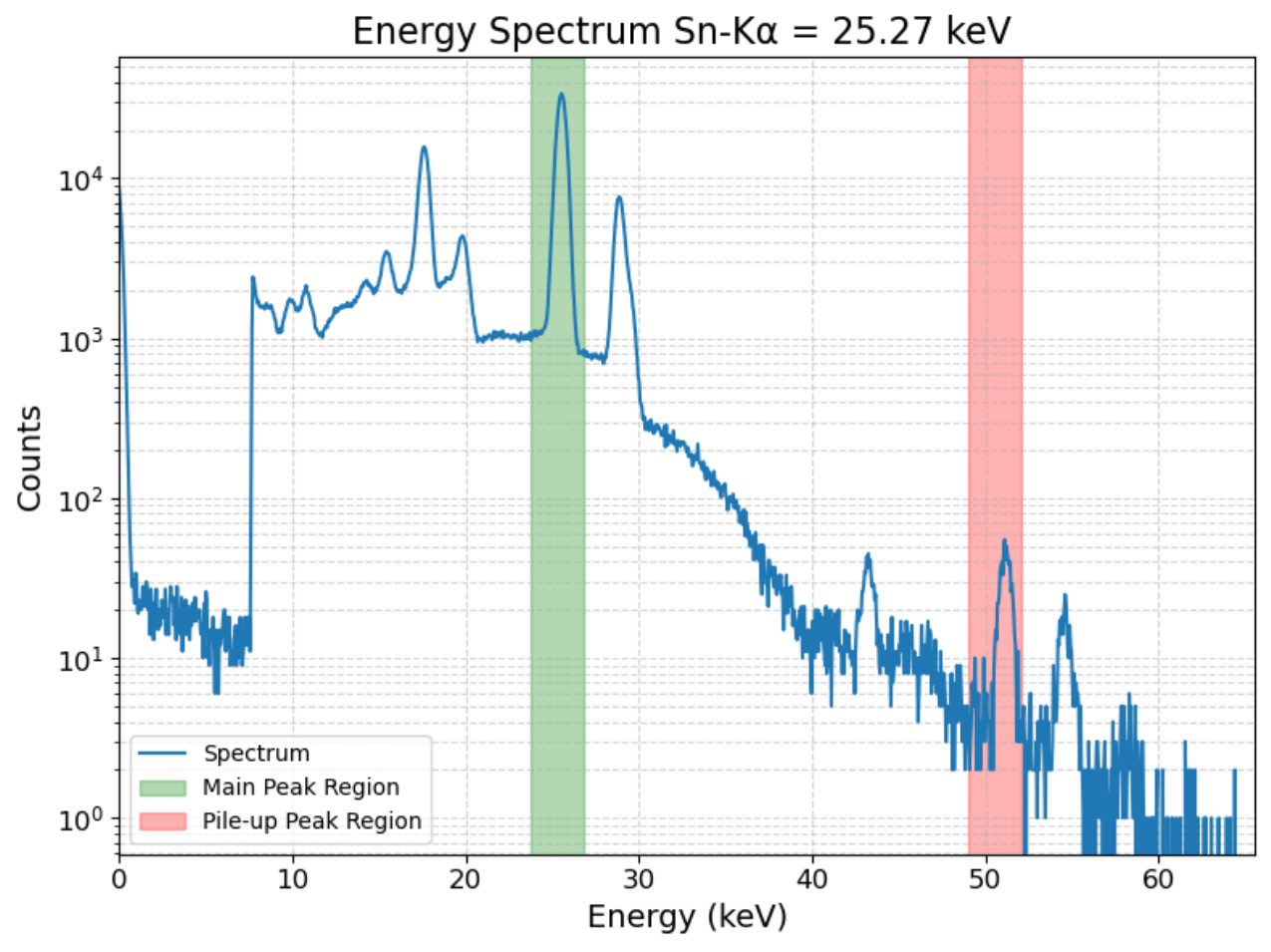}
        \caption{}
        \label{fig:sn_spectrum}
    \end{subfigure}
    \hfill
    \begin{subfigure}[b]{0.49\textwidth}
        \includegraphics[width=\textwidth]{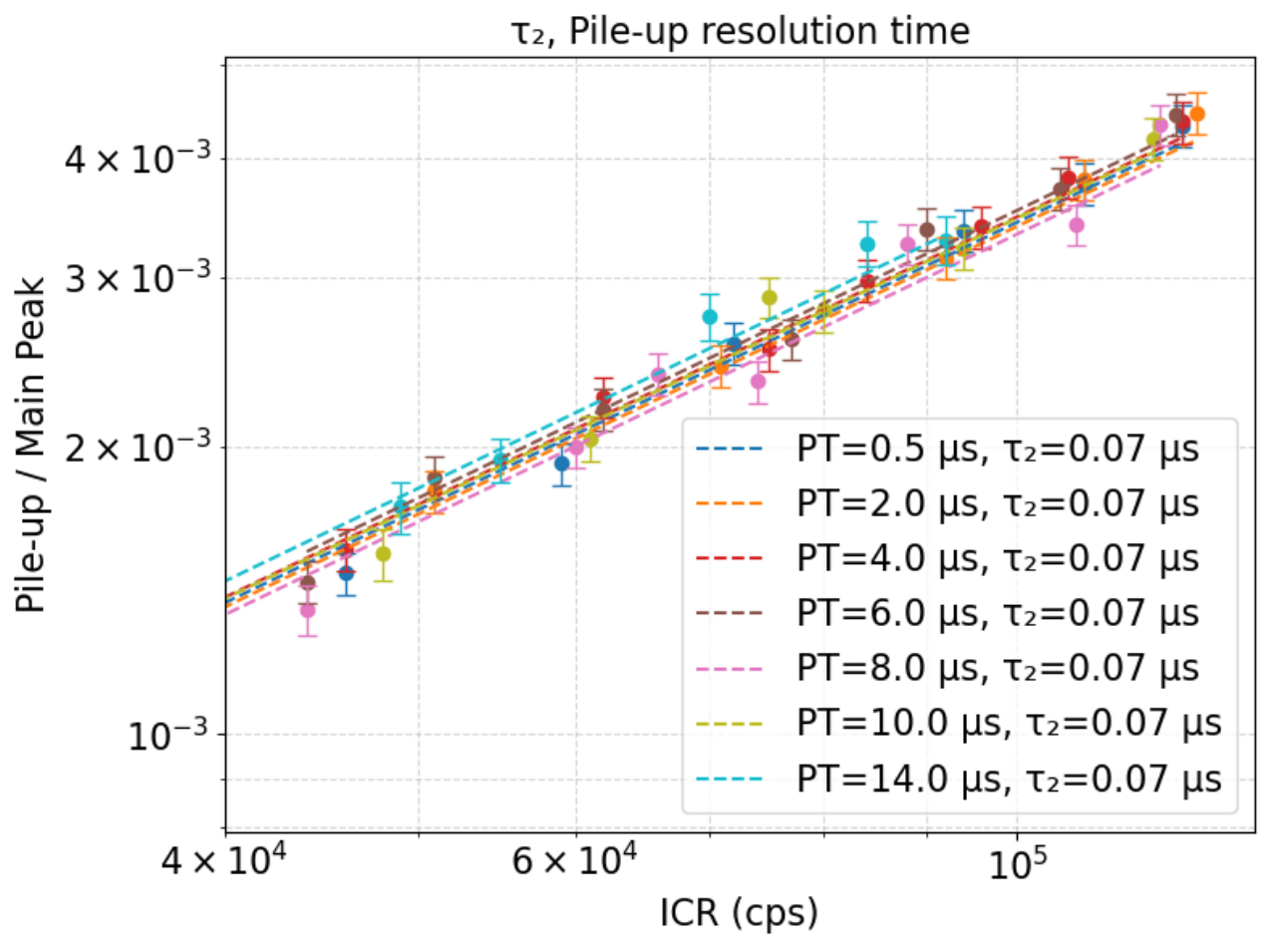}
        \caption{}
        \label{fig:sn_tau2}
    \end{subfigure}
    \begin{subfigure}[b]{0.55\textwidth}
        \includegraphics[width=\textwidth]{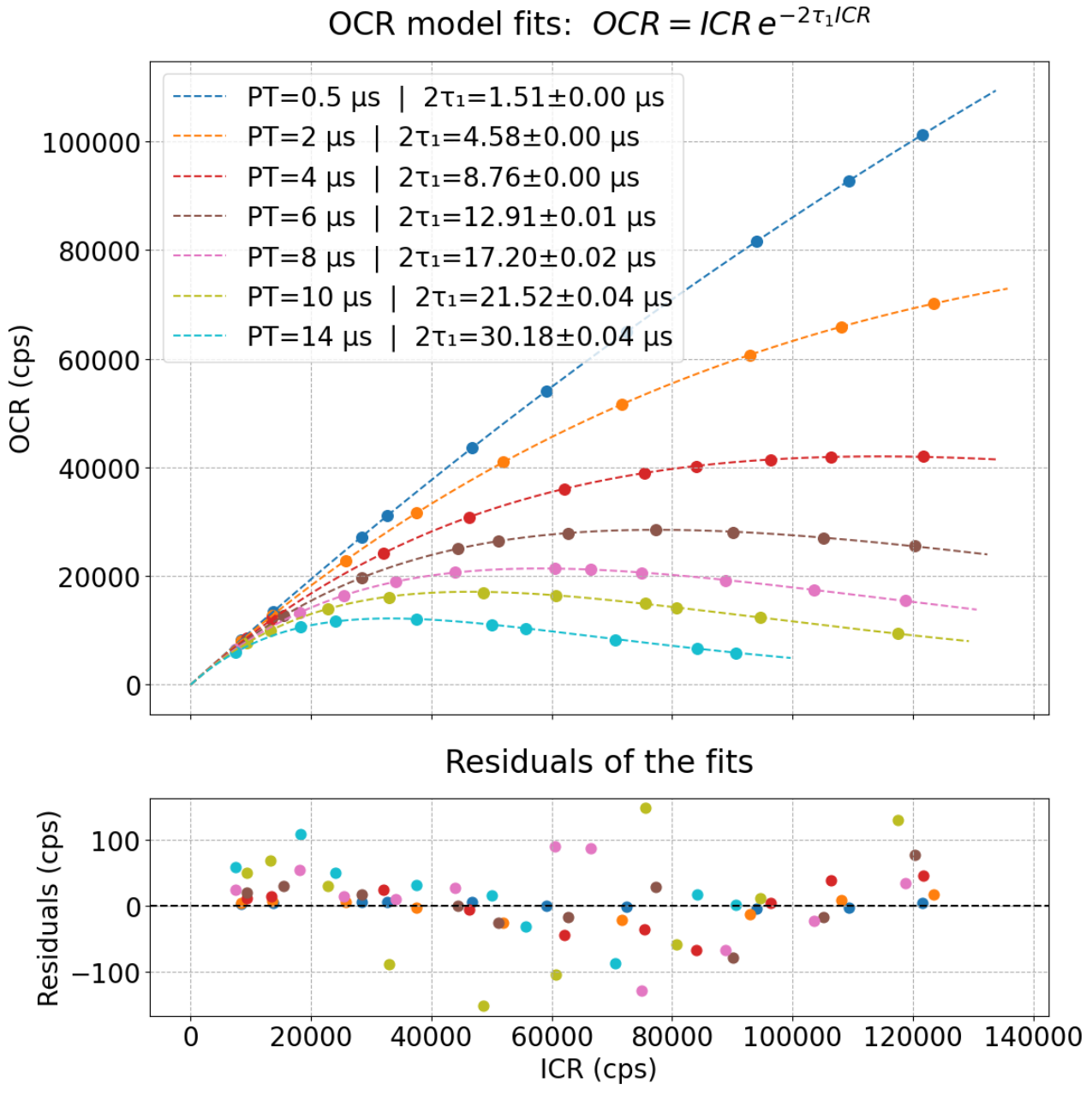}
        \caption{}
        \label{fig:OCRmodelfit}
    \end{subfigure}
    \caption{
(a) Energy spectrum acquired from a Sn target foil with characteristic fluorescence line at 25.27~keV. The main peak and pile-up peak regions are highlighted, and the Gaussian integral of the two regions is used to compute the pile-up ratio.  
(b) Pile-up ratio as a function of ICR for increasing  peaking time values. 
(c) OCR as a function of ICR at 25.27~keV w.r.t Peaking times, plotted along with fit residuals. 
}
    \label{fig:pileup_analysis}
\end{figure}

\subsubsection{Dead time and pile-up studies}
\label{sec:deadtimeandpileup}
We characterized the detector's performance under high-flux conditions by establishing a relationship between the input count rate (ICR) and the output count rate (OCR). Since incoming photons follow a Poisson process, the time interval between successive arrivals is exponentially distributed. For a spectroscopic detector, an event must be sufficiently isolated in time to be processed as an individual pulse; if another photon arrives too close in time, the two pulses sum together (pile-up), hence distorting the shape of the energy spectrum~\cite{blaj2017optimal}.

Under these assumptions, the probability that a photon is correctly resolved is given by the paralyzable model
\begin{equation}
    OCR = ICR \, e^{-2ICR \, \tau_1},
    \label{eq:paralyzable_model_eq}
\end{equation}
where $\tau_1$ is the detector’s effective processing time. This expression follows directly from the Poisson probability of observing no additional photons within the processing window of duration 2$\tau_1$. The parameter $\tau_1$ embeds the properties of the slow channel of the pulse processor.  This simple model was used to fit $\tau_1$ from our measured OCR and ICR data. Fitted values are in good agreement with the pulse processor settings.

The time constant of the pulse processor fast channel can also be extracted from the data. At high ICR, pile-up peaks appear at twice the energy of the main fluorescence line, i.e., the fast channel is unable to resolve two or more photons arriving as a bunch within a short time interval, as seen in Figure~\ref{fig:sn_spectrum}. According to Poisson statistics, the output count rate of pileup events involving 2 photons is given by the expression :
\begin{equation}
ICR^2 \times \tau_2 \times e^{-2 \times ICR \times \tau_1}  
\end{equation}

By estimating the output count rate ratio $ R_{\text{pileup}} $ between the pileup events involving 2 photons and  the count rate of single events, we could estimate the time constant of the pulse processor fast channel: electronics estimated the ratio of the fitted the main peak and the corresponding pile-up peak with Gaussian functions, and defined the pile-up ratio as
\begin{equation}
    R_{\text{pileup}} = \frac{ICR^2  \tau_2\, e^{-2 \times ICR \times \tau_1}}{ICR \, e^{-2\times ICR \times \tau_1}}\end{equation}
    \hfill
    \begin{equation}
        ICR \, \tau_2  \approx\frac{A_{\text{pileup}}}{A_{\text{main}}}
\end{equation}
where $A_{\text{pileup}}$ $(A_{\text{main}})$ are the Gaussian-fitted pileup (main) peak areas.

From our measurements, we obtained a $\tau_2 \approx 70~\text{ns}$ for all PTs, see Figure~\ref{fig:sn_tau2}.
Extended models that include additional parameters such as a reset duration term $\tau_3$ of CSA have been proposed in the literature~\cite{Bordessoule2019}. These models account for the effect of the preamplifier reset dead time at very high flux or higher photon energy. In our measurements, no significant saturation behaviour was observed. The data remained in a regime in which the response could be well described by a simple paralyzable model, without requiring an additional saturation parameter, as shown in Figure~\ref{fig:OCRmodelfit}. We initially tested models that included the CSA reset duration. however, the data are well described by a simple paralyzable model, indicating that the CSA reset duration had a negligible effect on the detector's OCR in the investigated regime.


\subsection{Surface scanning: Micro spot beam}
\label{sec:microsource_tests}
\begin{figure}[htb!]
\centering
  \includegraphics[height=4.5cm, width =\textwidth]{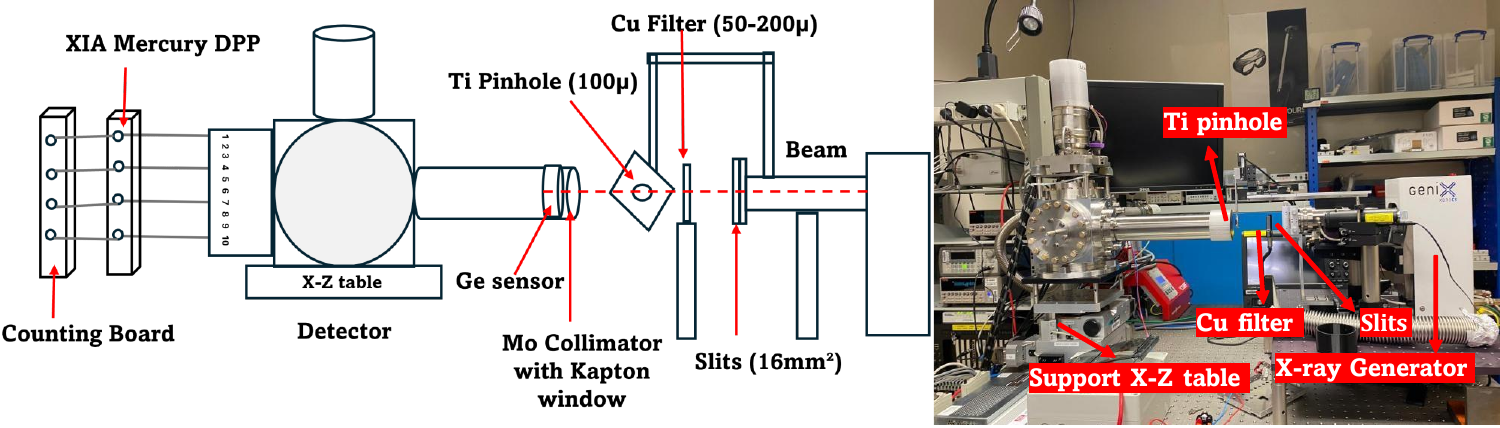}
\caption{Schematic of experimental setup for micro-source surface scans.}
\label{fig:microsource_setup}
\end{figure}

To investigate the spatial response, we performed surface scans using the microspot X-ray beam from the generator installed at the ESRF’s detector unit laboratory. The experimental setup, schematically illustrated in Figure~\ref{fig:microsource_setup}, consisted of a copper anode X-ray source delivering an 8.04~keV beam. The beam was collimated using motorized slits, followed by a 100~$\mu$m-diameter titanium pinhole to achieve a well-defined spot on the detector surface. Beam flux was tuned by inserting copper filters of varying thickness in the beam path.\\
\par
The detector was mounted on a motorized X-Z stage for automated scanning and was biased at +70~V throughout the measurements. Four pixels were read out simultaneously via an XIA-Mercury digital pulse processor (DPP), allowing us to use the TTL outputs from the user-defined region of interest (ROI) in the energy spectrum. It enabled easy integration with external counting systems which count only events within the Cu~K$\alpha$ peak ROI at 8.04~keV, see Figure~\ref{fig:microsource_setup}. We performed the mapping of sensor surface with a scan step size of 50~$\mu$m and an exposure time of 1~s per position, covering three to four pixels per scan area. The main scan parameters are summarized using Table~\ref{tab:scan_parameters}.\\
\begin{table}[htb!]
\centering
\caption{Parameters for surface scans with the micro-focus X-ray source.}
\vspace{0.5cm}
\label{tab:scan_parameters}
\begin{tabular}{ll}
\hline
\textbf{Parameter} & \textbf{Value } \\ \hline
X-ray source & Micro-focus Cu anode (ESRF CB182A) \\
Beam energy & 8.04~keV (Cu-K${_\alpha}$) \\
Beam spot size & $\sim$100~$\mu$m \\
Collimation & Two motorized slit pairs + Ti pinhole (100~$\mu$m) \\
Flux attenuation & Cu filters: 50, 100, 150, 200~$\mu$m \\
Bias voltage & +70 - 150~V \\
Readout electronics & XIA-Mercury DSP + TTL counter \\
Calibration & \textsuperscript{55}Fe offline source (Mn-K$\alpha$ at 5.9~keV) \\
Scan step size & 50~$\mu$m \\
Exposure time per step & 1~s \\
Number of scans & 4-5 (covering 3--4 pixels/scan) \\
\hline
\end{tabular}
\end{table}


\begin{figure}[h!]
  \centering
  \begin{subfigure}[t]{0.48\textwidth}
    \centering
    \includegraphics[height=6cm,keepaspectratio]{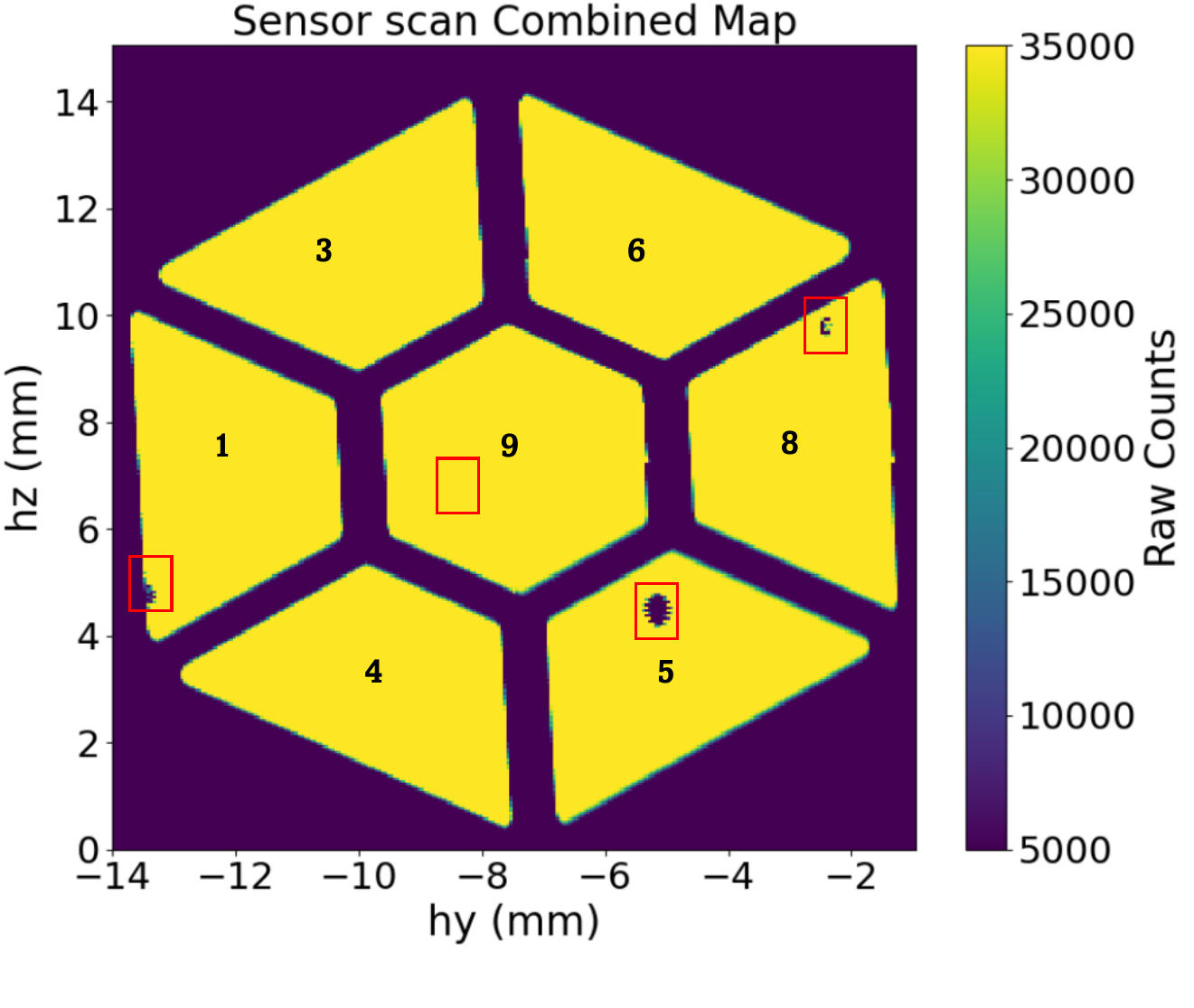}
    \caption{}
    \label{Fig:geom_full}
  \end{subfigure}
  \hfill
  \begin{subfigure}[t]{0.48\textwidth}
    \centering
    \raisebox{0.3cm}{\includegraphics[height=6cm,keepaspectratio]{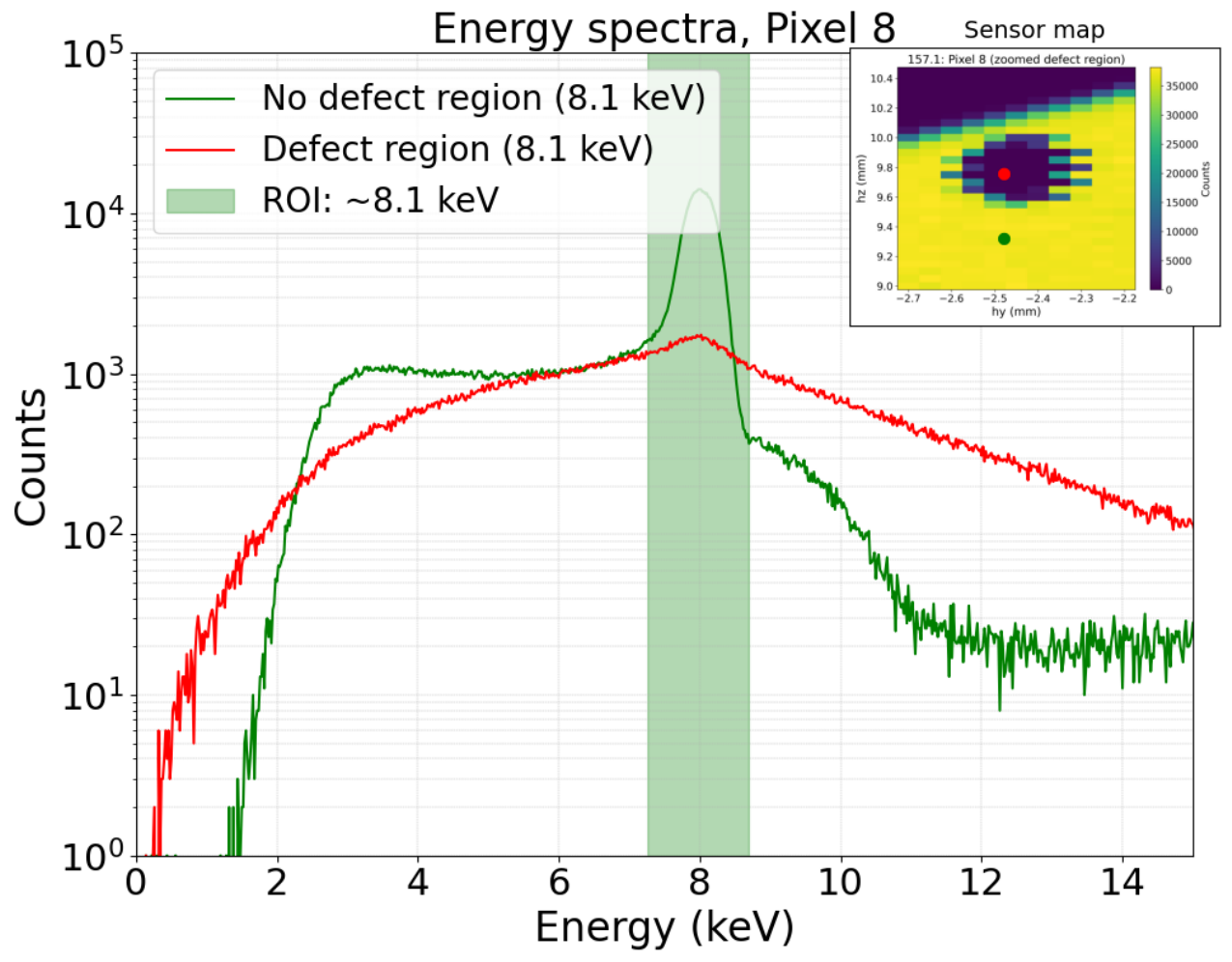}
    }
    \caption{}
    \label{fig:energyspectra_goodvsbad}
  \end{subfigure}

  \caption{(a) Fully reconstructed sensor image, obtained with the combination of several pixel scans. (b) Energy spectra recorded at 8.04~keV from a defect region and a normal active area of the same pixel. The defect spectrum shows a significant loss in total counts and peak height compared to the normal region.}
\end{figure}

\begin{figure}[htb!]
\centering
\hfill
\begin{subfigure}[b]{\linewidth}
  \centering
  \includegraphics[width=0.8\textwidth, keepaspectratio]{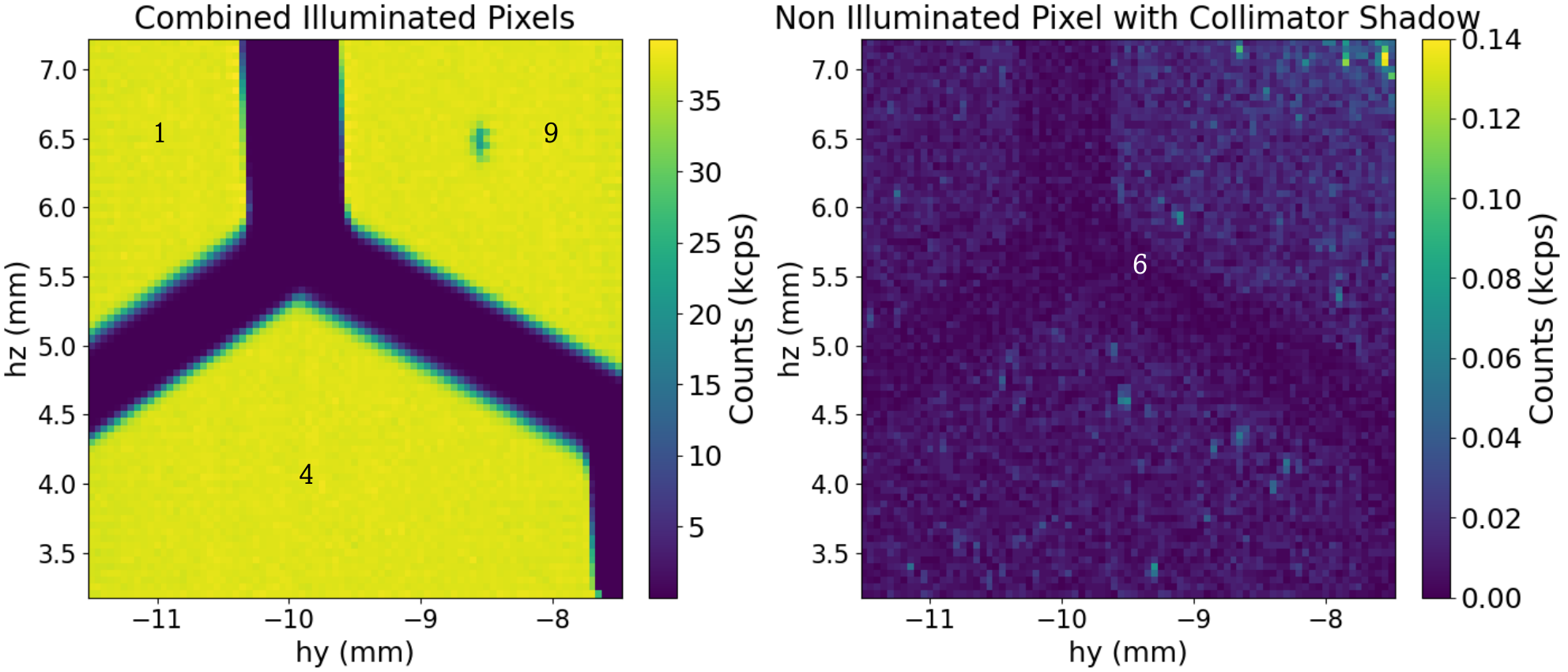}
  \caption{}
  \label{Fig:crosstalk}
\end{subfigure}
\vspace{1em}
\begin{subfigure}[b]{0.80\linewidth}
  \centering
  \includegraphics[width=0.85\textwidth, keepaspectratio]{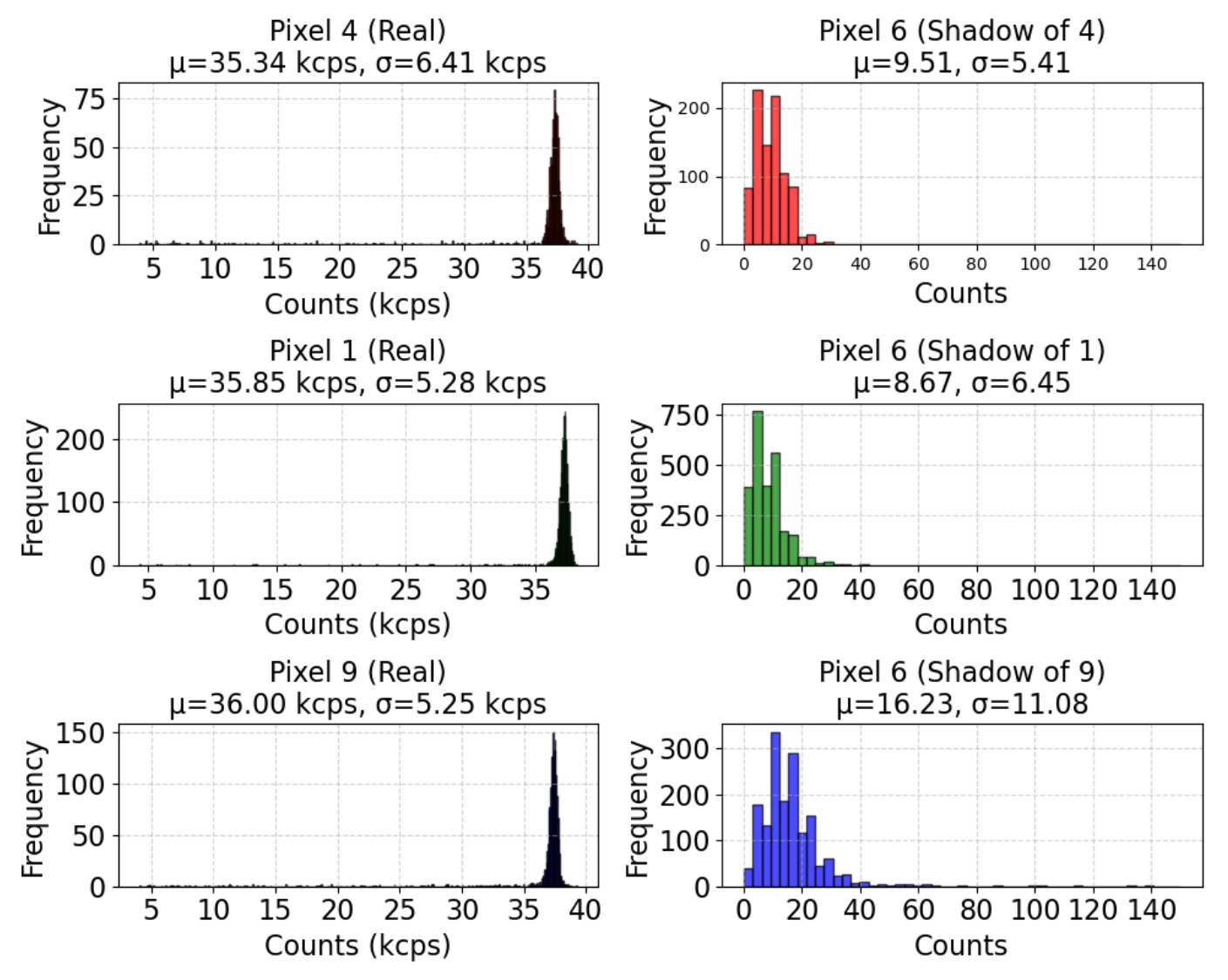}
  \caption{}
  \label{Fig:crosstalk_histos}
\end{subfigure}

\caption{(a) The combined counts from the three directly illuminated pixels, showing the collimator edge (left) and a non-illuminated pixel that shows a faint shadowed image of the illuminated pixels. (b) Histograms illustrating the observed  count rate in the illuminated pixels and the non-illuminated ones.}
\label{fig:illumshadow}
\end{figure}
\par
The reconstructed surface map obtained by combining individual scans is shown in Figure~\ref{Fig:geom_full}.
To create a  spatial response map of the entire sensor, data from multiple, smaller raster scans were combined into a single heatmap. Since individual scans covered only a few pixels or a specific region of the detector, this "patchwork" approach was essential for visualizing the performance across the full sensor surface.
The reconstruction process was first performed by compiling all the individual scans, along with the motor coordinates (hy, hz) and the corresponding signal counts for each active detector pixel.
We then aggregated this information into a single dataset to create a 2-D histogram, which served as the canvas for the final image. A uniform bin width of 50~$\mu$m was chosen to match the high-resolution scan step size. The ranges of the histogram's x and y axes were set to cover the full extent of all motor positions recorded across all scans.
The aggregated data points were then binned into the 2D histogram. The value of each bin in the final map was determined by the total number of counts recorded at that specific (x, y) spatial coordinate, summed across all contributing pixels from all scans.
As a result, we confirmed that the pixel boundaries closely match the molybdenum collimator apertures, confirming excellent alignment between the collimator and the sensor.
\par
Four distinct regions of reduced sensitivity were detected (highlighted in red), see Figure~\ref{Fig:geom_full}, corresponding to the defective area in the germanium sensor. When the X-ray beam  (count rates of 30-35~kcps) illuminated these areas, the leakage current increased from its nominal value of 1~pA to approximately 100~pA.
\par
Except these defective regions, the detector showed a uniform behavior across the active area.
Although the majority of pixels had stable performance, Small variations were observed in  count-rate statistics. 
Measured count rate dispersions ranged from 1\% (pixel~5) to 17\% (pixel~9). The latter, corresponding to the central pixels, showed this response because we combined three individual scans to reproduce the full pixel-count rate distribution.

To further investigate the defective areas, energy spectra were acquired at positions within (black regions) and outside (yellow region) each defect. A typical example is shown in Figure~\ref{fig:energyspectra_goodvsbad}. The spectrum from within the defect shows reduced total counts and a suppressed Cu-K$_{\alpha}$ peak, whereas the normal region shows the expected photopeak with nominal intensity. These results confirm that the observed low-response zones in the surface map correspond to true inactive regions in the sensor.
\subsubsection{Crosstalk between pixels}
\label{sec:crosstalk_analysis}
To study the crosstalk between pixels, we positioned the beam to cover the interpixel spacing, i.e., the collimator edge. In this position, three out of four pixels (max.) were directly illuminated, while the rest of the neighboring pixels received no direct exposure. In this measurement, for each scan, one non-illuminated pixel was connected to the readout electronics.
Crosstalk analysis, illustrated in Figure~\ref{Fig:crosstalk}, revealed that when one pixel was scanned, a neighboring pixel not directly illuminated registered a faint shadow of the collimator.  
\par
To quantify crosstalks, we defined an illumination mask based on the bright regions in the illuminated pixels (above 10\% of the maximum counts). For each illuminated pixel $i$ and each non-illuminated pixel $j$, we calculated:
\[
\mathrm{XT}_{ij} = \frac{\langle C^{(j)}\rangle}{\langle C^{(i)}\rangle} \times 100\%
\]
Where $\langle C^{(i)}\rangle$ is the average counts in the illuminated pixel inside the mask, and $\langle C^{(j)}\rangle$ is the average counts from the corresponding non-illuminated pixel over the same positions.
The shadow-image shape matches the collimator edge, suggesting the effect arises from electronics channel coupling.
The histograms comparing the counts in illuminated and non-illuminated pixels are shown in
Figure~\ref{Fig:crosstalk_histos}. 
The measured crosstalk ranged from 0.024\% to 0.045\%, well below the 0.2\% specification provided by the ASIC manufacturer XGlab.

\subsubsection{Rise Time Scanning}
\begin{figure}[h!]
    \centering
    \begin{subfigure}[b]{0.25\textwidth}
        \centering
        \includegraphics[height=5cm]{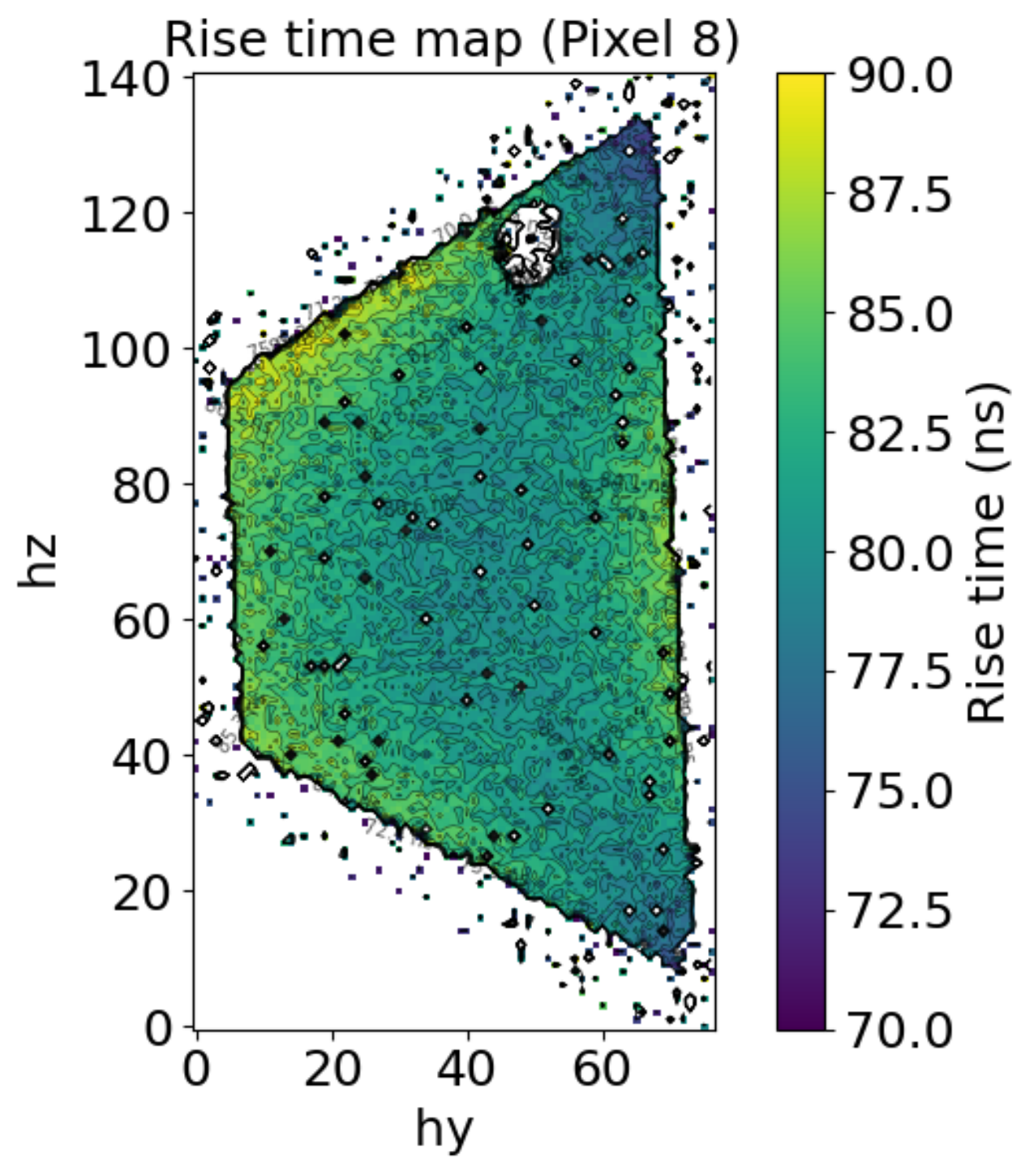}
        \caption{}
    \end{subfigure}\hfill
    \begin{subfigure}[b]{0.36\textwidth}
        \centering
        \includegraphics[height=5cm]{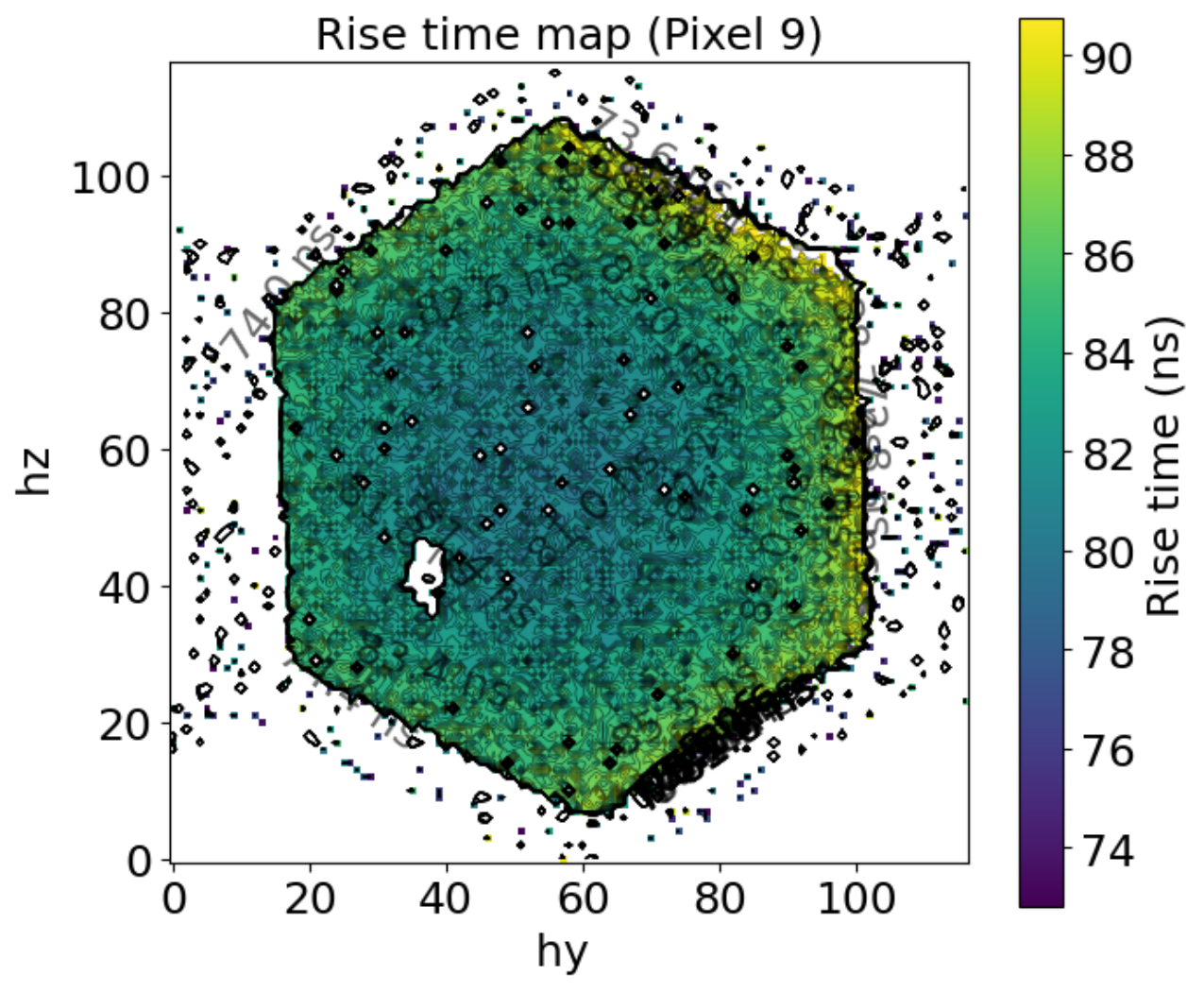}
        \caption{}
    \end{subfigure}\hfill
    \begin{subfigure}[b]{0.31\textwidth}
        \centering
        \includegraphics[height=5cm]{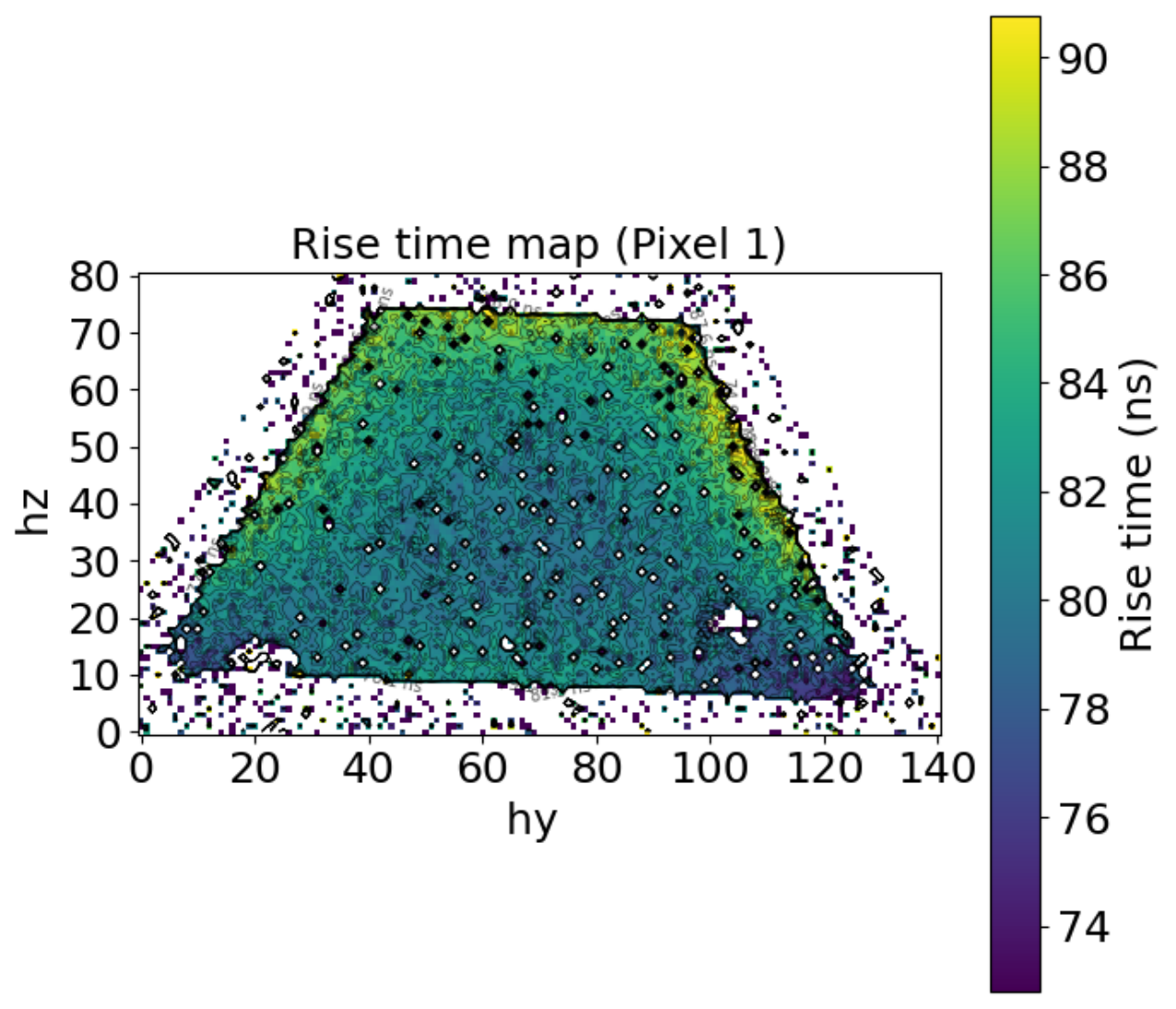}
        \caption{}
    \end{subfigure}
    \caption{Rise time scans for pixel~8~(a), pixel~9~(b), and pixel~1~(c). The region affected by defects shows complete charge loss (shown in white, inside the pixel active area).}
    \label{Fig:Risetime_map}
\end{figure}

We evaluated the timing performance by performing Rise-time scans using a digital oscilloscope coupled to the detector output. The oscilloscope was configured with a horizontal time base of 200 ns/div and a vertical sensitivity of 20–40 mV/div (depending on signal amplitude). Data were acquired as fall-time histograms, with each mesh step corresponding to a 4~s integration time. The spatial scan was performed in a 2D mesh pattern, using a 50 $\mu$m step size. We measured the signal risetime distribution for three pixels, as shown in Figure~\ref{Fig:Risetime_map}. The average risetime was approximately 80~ns at +70 V, consistent across the pixel surface, except for some edge effects, visible along the edges of three pixels. The measured risetime is consistent with the simulated drift time of charge carriers in the sensor (50-70 ns at +70V).


\section{First On-beam tests}
\label{sec:beamtests}
The first synchrotron measurements with our 1$^{st}$ prototype detector were performed at the BM05 beamline in the ESRF-EBS facility~\cite{BM05beamline}. The BM05 beamline provides an experimental hutch with a focused monochromatic X-ray beam, achieving a spot size of approximately 35~$\mu$m for detector characterization and follow-up spectroscopy tests.

\begin{figure}[htb!]
\centering
  \includegraphics[height=9.0cm, width =0.85\textwidth]{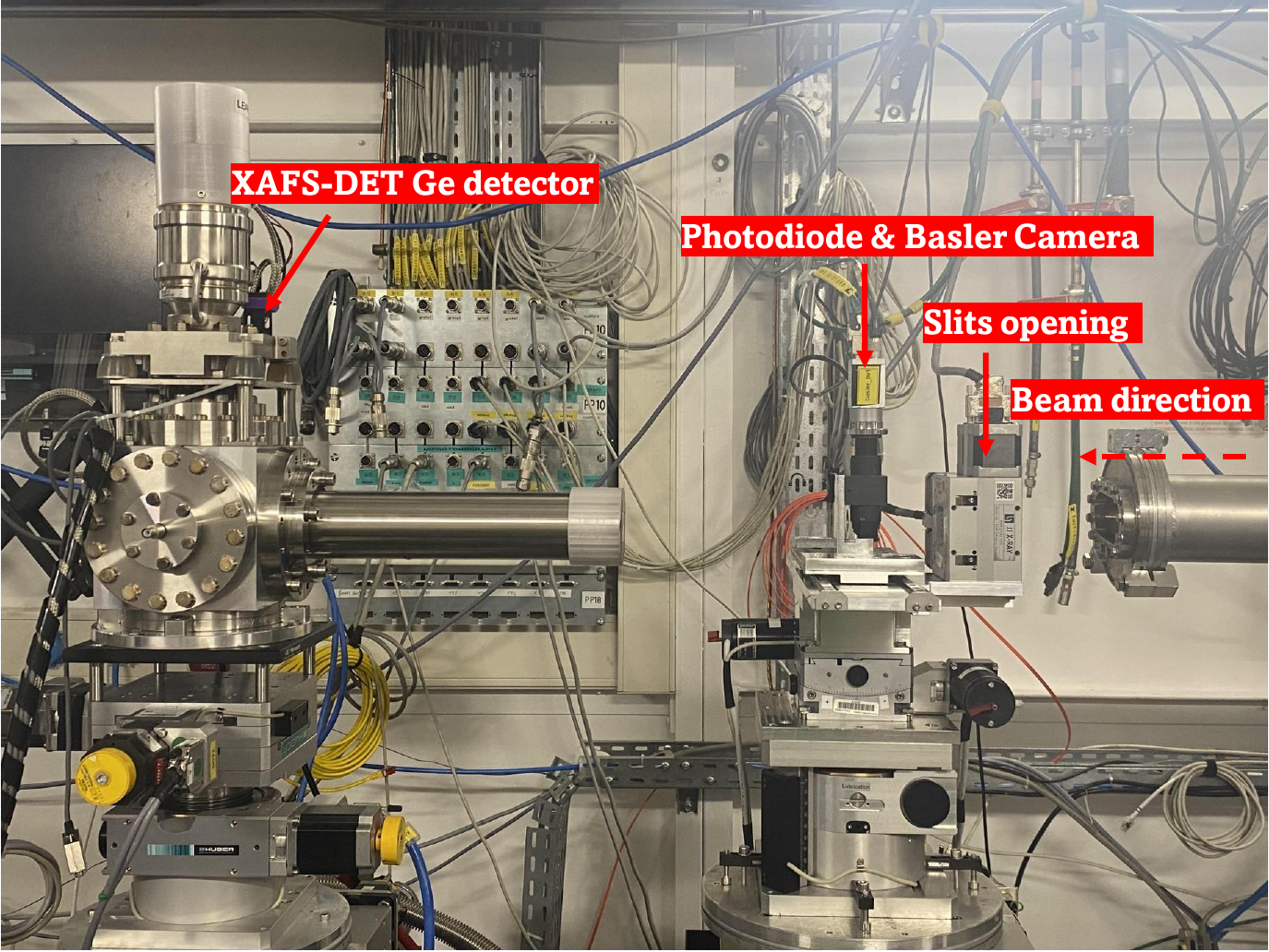}
\caption{Schematic of Experimental setup for beam tests. A photodiode and a Basler camera were used as diagnostic tools for measuring incident beam intensity and for beam alignment.}
\label{Fig:beamtest_setup}
\end{figure}

The detector was mounted on a two-axis motorized stage for precise alignment with the incident beam. In the same hutch, a dedicated diagnostics stage was equipped with a standard photodiode for monitoring beam intensity and a Basler camera to visualize the beam spot size. A schematic illustration of the final setup is presented in  Figure~\ref{Fig:beamtest_setup}.


\begin{table}[htb!]
\centering
\caption{Summary of experimental parameters for the on-beam tests at ESRF BM05.}
\hfill
\label{tab:beamtest_params}
\begin{tabularx}{\textwidth}{lX} 
\toprule
\textbf{Parameter} & \textbf{Value} \\
\midrule
\multicolumn{2}{c}{\textbf{Beamline and Beam Parameters}} \\
\midrule
Beamline & ESRF BM05 \\
Beam Type & Monochromatic, Focused \\
Beam Spot Size & \textasciitilde 35 $\mu$m \\
Photon Energies & 20, 30, 40, and 50~keV \\
Photon Flux & Tunable (controlled via slits and filters) \\
Beam Diagnostics & Calibrated Photodiode, Basler Camera, LaBr3 Scintillator+PMT \\
\midrule
\multicolumn{2}{c}{\textbf{Scan Parameters}} \\
\midrule
Bias Voltages & 70 V, 150 V, 250 V \\
Low-Resolution Scan Step & 250~$\mu$m \\
High-Resolution Scan Step & 25~$\mu$m \\
 \bottomrule
\end{tabularx}
\end{table}
\begin{figure}[htb!]
    \centering

    \begin{subfigure}{0.55\textwidth}
        \centering
        \includegraphics[width=\linewidth]{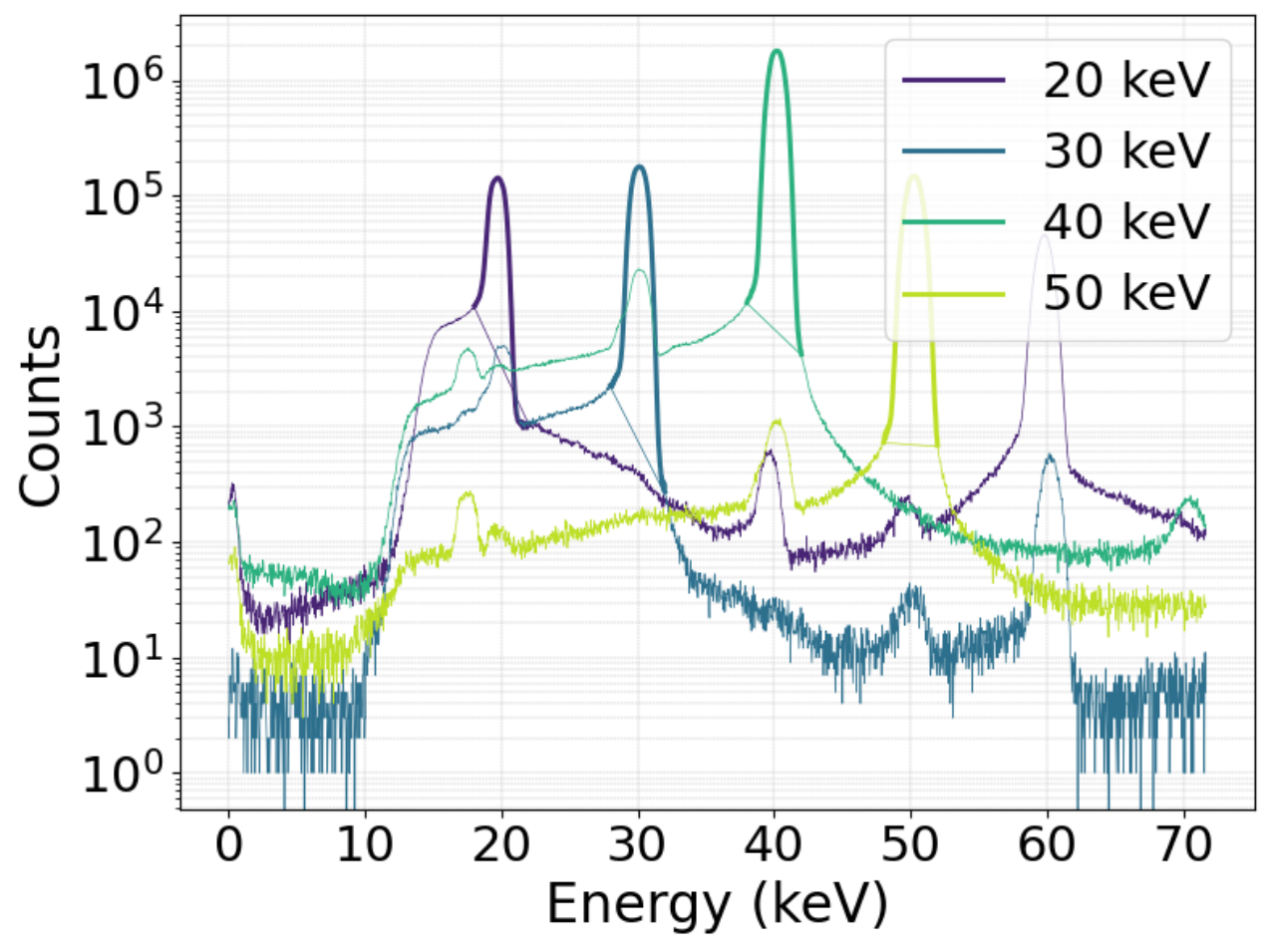}
        \caption{}
        \label{fig:directenergy}
    \end{subfigure}
  
    \begin{subfigure}{0.49\textwidth}
        \centering
        \includegraphics[width=\linewidth]{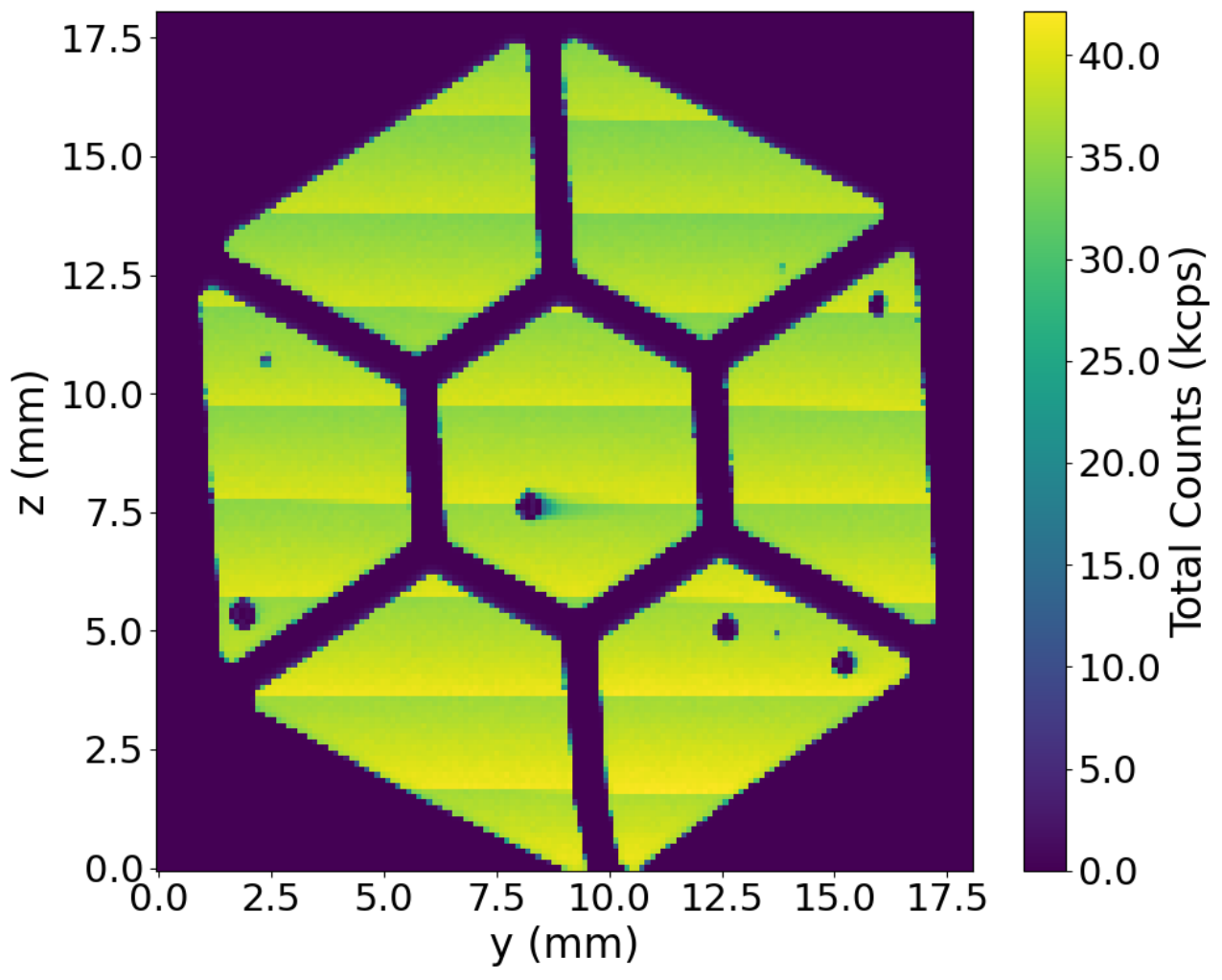}
        \caption{}
        \label{sensor_image_directenergy}
    \end{subfigure}
    \hfill
    \begin{subfigure}{0.49\textwidth}
        \centering
        \includegraphics[width=\linewidth]{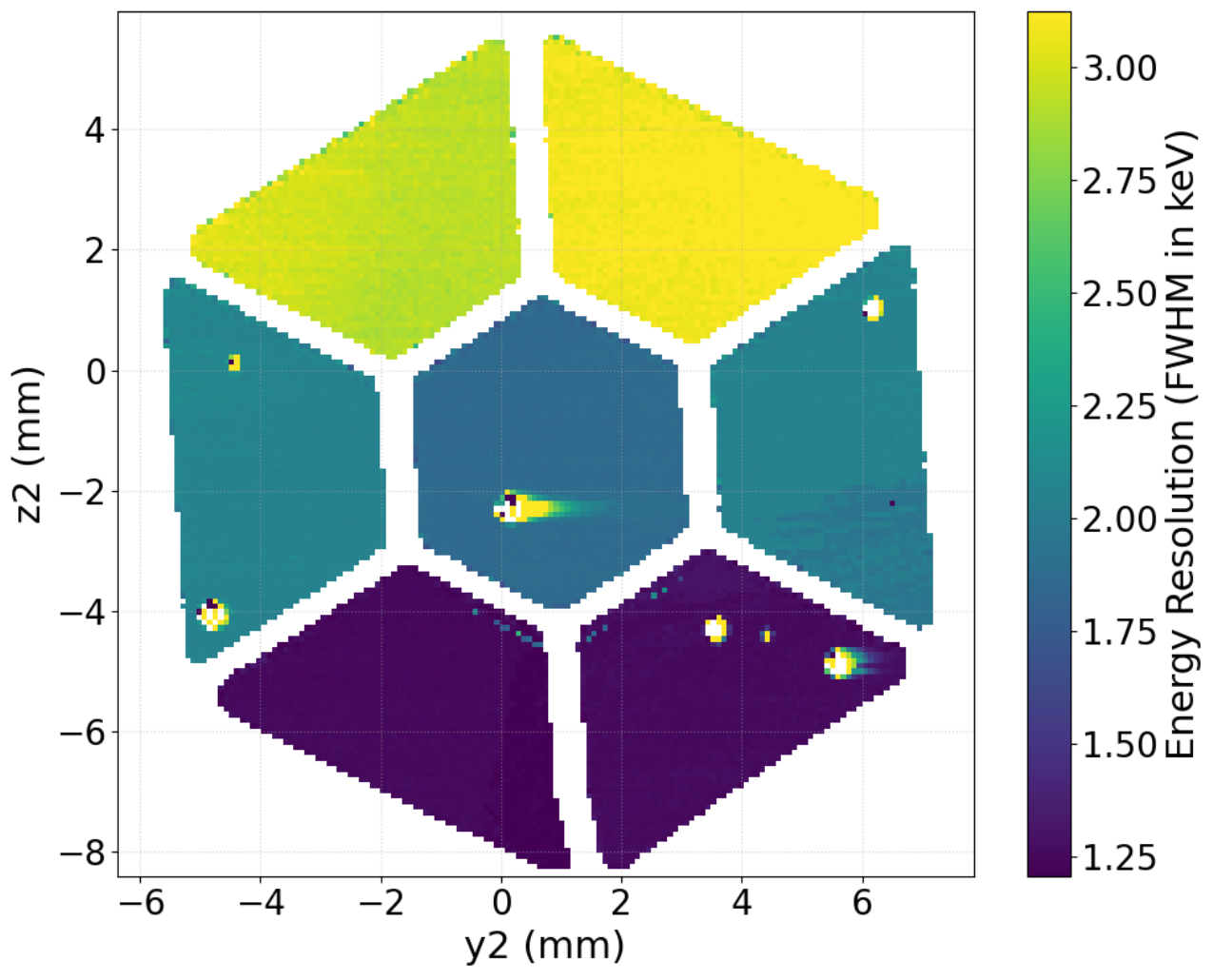}
        \caption{}
        \label{fwhm_directenergy}
    \end{subfigure}

    \caption{(a) Direct incident energy beam respectively at 20, 30, 40 and 50~keV. (b) A heatmap of the total counts recorded across the sensor during a 40~keV scan. The dark spots correspond to sensor defects. (c) Map of the corresponding energy resolution (FWHM) for a 40~keV scan. The bright spots, indicating poor resolution, are the defects identified in Figure~\ref{sensor_image_directenergy}.}
\end{figure}

\subsection{Performance at Monochromatic Energies}
\label{sec:performabce_directeenrgy}
During the first beam tests, we acquired several two-dimensional spatial response maps as well as the energy spectrum, (see Figure~\ref{fig:directenergy}), for each scan point to assess detector performance (energy resolution at 40~keV, shown in Figure~\ref{fwhm_directenergy}), and investigate the localized defects previously identified in laboratory micro source beam scans.
We performed high-resolution X-ray scans by exposing the sensor surface, and specifically the defect area, to high-energy photons and by exploring their behavior under different detector biases. The idea of modeling the suppression to estimate defect depth emerged later, followed by the curiosity about the origin of the defects. This led to the development of a simple yet effective model that enabled us to estimate the depth of the defects from the suppression data of each scan.
The beam and the scan parameters are tabulated in Table~\ref{tab:beamtest_params}.
\par
\begin{figure}
      \centering
      \includegraphics[width=\textwidth]{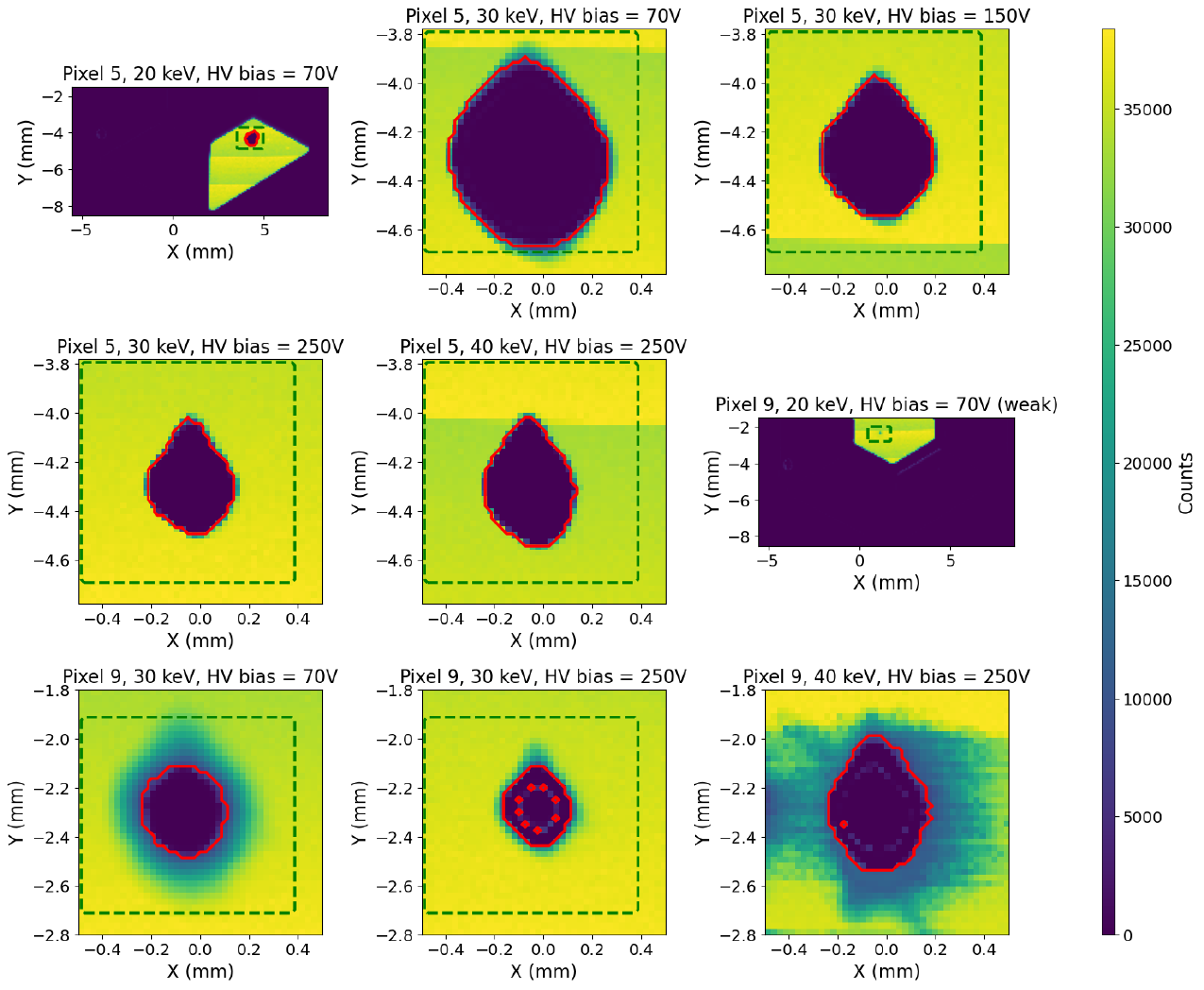}
    \caption{2D maps with  regions of interest (ROI),
Each subplot shows the spatial count distribution for pixel 5 and pixel 9, at a given X-ray energy, and applied bias.
Green-dashed contours represent the region of interest (ROI), whereas red contours showcase the suppressed (defective) areas.}
    \label{fig:def:ROI}
  \end{figure}

Two types of scans were performed. First, low-resolution raster scans (250~$\mu$m step size) across the entire sensor area at the three aforementioned energies to generate a comprehensive performance map. These scans showed several defective regions, as evidenced by pixels 1, 5, 8, and 9, which exhibited significant signal loss, as shown in Figure~\ref{sensor_image_directenergy}. Following this, we acquired high-resolution scans (25~$\mu$m step size), focusing mainly on the defective areas of pixel 5 and 9. These targeted scans were conducted at three bias voltages to obtain the necessary data for a quantitative analysis of the defect's behaviour, as shown in Figure~\ref{fig:def:ROI}.
\subsection{Defect depth estimation} 
\label{sec:defectdepth_analysis}
With this analysis, we estimated the physical depth of the defects based on measured signal suppression. We used two-dimensional heatmaps of the defect region to define an ROI containing the defect. The severity of the defect is evaluated using the signal/count suppression ratio $S$, defined as the ratio of the average counts in the defective region $I_{\text{defect}}$ to the maximum counts in the good region $I_{\text{norm}}$, as shown using eq.~\ref{eq:suppresssionratio}. \\

\begin{equation}
S = \frac{I_{\text{defect}}}{I_{\text{norm}}}
\label{eq:suppresssionratio}
\end{equation}


\par
The suppression of the X-ray signal in the presence of a defect arises from two physical effects: absorption of the incident photon as it travels deeper into the material, and a reduction in charge-collection efficiency due to carrier trapping. We model this phenomenon as:
\par
\begin{equation}
S(d) = e^{-\mu(E) d} \cdot \left(1 - e^{-d / \lambda_{\text{eff}}} \right)
\label{eq:suppression_model}
\end{equation}
In this expression, $\mu$ is the energy-dependent linear attenuation coefficient of germanium (obtained from xraylib~\cite{brunetti2004library}), while $\lambda_{\text{eff}}$ represents the mean drift length, which accounts for the average distance an electron-hole pair can travel before being trapped. Since a stronger field improves the charge  transport and therefore the charge collection, we model $\lambda_{\text{eff}}$ as a voltage-dependent parameter, increasing with the applied bias $V$:
\begin{equation}
\lambda_{\text{eff}}(V) = \lambda_0 \cdot (1 + \alpha V)
\end{equation}

In this formulation, the two parameters ($\lambda_0$ and $\alpha$) are effective parameters that capture the voltage-dependent charge-collection behavior of the germanium bulk. Physically, $\lambda_0$ represents the intrinsic mean drift length of charge carriers at zero applied bias, dictated by intrinsic material quality such as sensor impurities or the density of trapping centers. 

The $\alpha$ quantifies how the effective collection depth increases with the electric field, showing a reduction in the trapping probability and an increase in the carrier drift velocity at higher bias. Both parameters were fitted globally across all bias and energy values by minimizing the deviation between the measured and predicted suppression ratios and the model, respectively. A similar approach has been shown in 
The best-fit parameters were found to be $\lambda_0 = 0.10$~mm and $\alpha = 0.0099$~V$^{-1}$.
For each (energy, bias) point, the defect depth $d$ was obtained by numerically inverting the suppression model given in eq.~\ref{eq:suppression_model}, by solving
\begin{equation}
f(d)= e^{-\mu(E)\, d} - e^{-(\mu(E)+1/\lambda_{\mathrm{eff}}(V))\, d} - S_{\mathrm{meas}} = 0
\label{eq:CCE_inveersion}
\end{equation}
\begin{table}[htb!]
\centering
\caption{Mean photon absorption depth $\langle x \rangle$ in germanium for photon energies between 8~keV and 50~keV, calculated using \texttt{xraylib}. The values include transmission losses through the Be and Kapton entrance window.}
\hfill
\label{tab:absorptiondepth}

\begin{tabular}{ccc}
\hline
\textbf{Photon Energy (keV)} & \textbf{Mean Absorption Depth $\langle x \rangle$ (mm)} & \textbf{Transmission $T_{\text{window}}$} \\
\hline
8  & 0.027 & 0.39 \\
20 & 0.044 & 0.93 \\
30 & 0.136 & 0.97 \\
40 & 0.303 & 0.98 \\
50 & 0.563 & 0.99 \\
\hline
\end{tabular}
\end{table}

\par
\begin{figure}
      \centering
      \includegraphics[height=10.5cm,width=\textwidth]{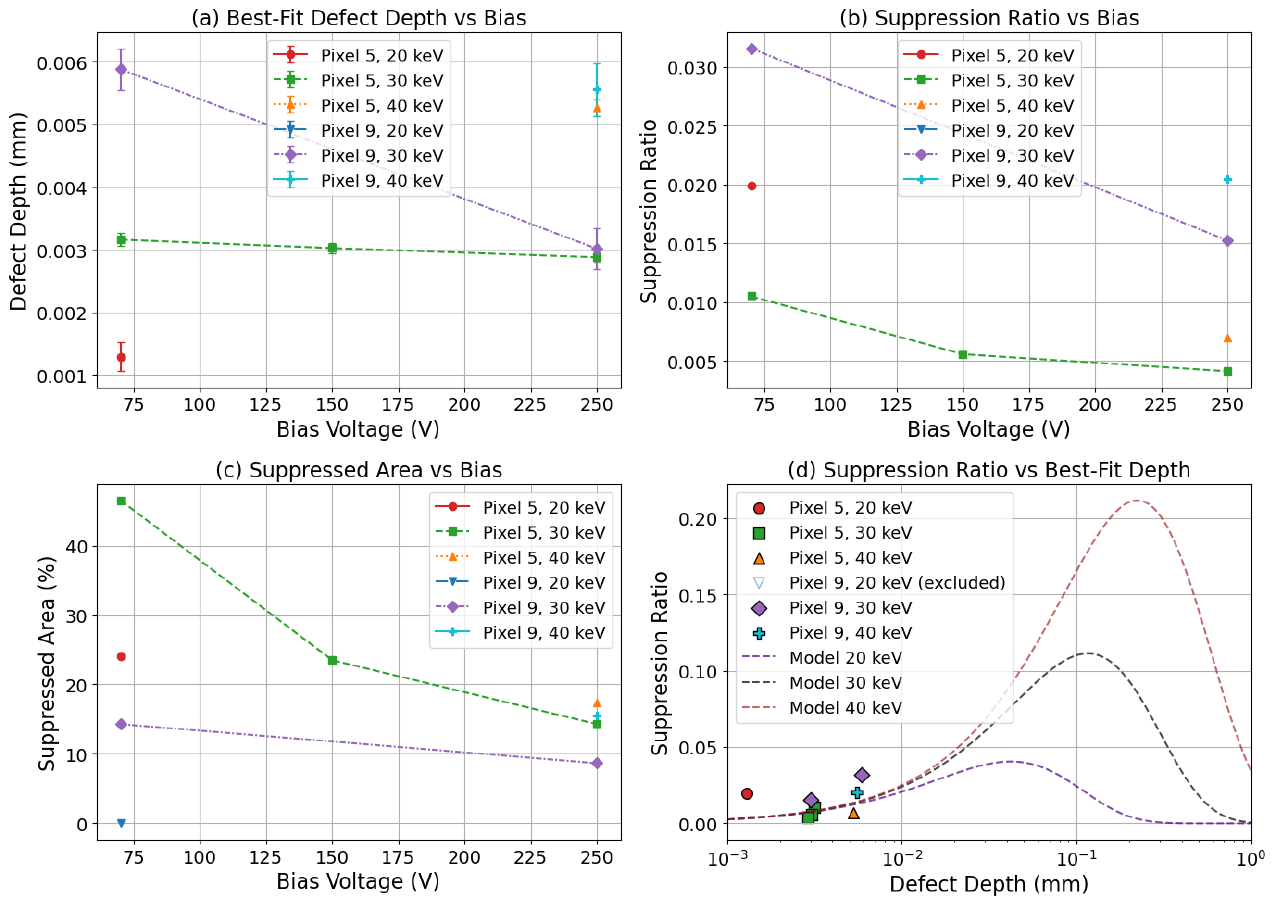}
    \caption{Comprehensive analysis of detector defects.
(a) Detector bias voltage vs. the best depth defect fit.
(b) Suppression ratios, showing that the defect in pixel 5 is more severe than in pixel 9.
(c) Suppressed area vs. bias voltage.
(d) Validation of the CCE model, showing all measured data points from different energies overlaid with theoretical model curves.}
    \label{fig:combined_analysis}
  \end{figure}

We summarize these analysis results using four subfigures of Figure~\ref{fig:combined_analysis}.\\
Figure~(a) shows the derived defect depths, indicating that the studied defects are shallow and exist closer to the sensor entrance surface, likely introduced during the sensor fabrication process (Please refer to Section 3.1 in ~\cite{Markamman_hpgesenosrgrowt}). \\
Figures~(b) and (c) show the measured suppression ratios and defect areas, respectively, showing that the defect in pixel 5 is severe as compared to the one observed in pixel 9. With increasing bias, both defects shows improved charge collection, resulting in less suppression and a smaller area. In our calculations, one data point was excluded from the final suppression ratio fit
\footnote{\label{scanpoint} The 20~keV scan of pixel 9 at 70~V was excluded because there was no clear defect contrast.}. 
\par

Finally, Figure~(d), shows the model curves from the CCE model using the fitted $\lambda_0$ and $\alpha$, overlaid with measured data points. The shape of the model curves can be interpreted from the competing product of photon attenuation and charge collection efficiency factor as shown in the model expression \ref{eq:suppression_model}. For shallow defects ($d \ll \lambda_{\text{eff}}$), the suppression rises linearly with depth since only a small fraction of created charge is lost. As $d$ approaches $\lambda_{\text{eff}}$, the charge collection saturates and $S(d)$ flattens, showing the transition from partial to nearly complete charge collection.
At large depths, the exponential attenuation term $e^{-\mu d}$ dominates, leading to a decrease at higher energies.
Increasing the detector bias effectively extends $\lambda_{\text{eff}}$, leading to shallower and less pronounced suppression curves, consistent with improved charge transport in stronger electric fields.\\

 \begin{table}[htb!]
\centering
\caption{Measured dimensions, suppression ratios, and fitted depths of the defects in pixels 5 and 9 at various photon energies and bias voltages. The dimensions are extracted from the 2D raster scans (see Figure~\ref{fig:def:ROI}), and the defect depths are obtained through inversion of the suppression model.}
\label{tab:defect_dimensions}
\hfill
\begin{tabular}{ccccccc}
\hline
\textbf{Pixel} & \textbf{Energy (keV)} & \textbf{Bias (V)} & \textbf{Width (mm)} & \textbf{Height (mm)} & \textbf{Suppression $S$} & \textbf{Depth (mm)} \\
\hline
5 & 20 & 70  & 0.628 & 0.718 & 0.020 & 0.0013 \\
5 & 30 & 70  & 0.624 & 0.750 & 0.010 & 0.0031 \\
5 & 30 & 150 & 0.425 & 0.550 & 0.005 & 0.0030 \\
5 & 30 & 250 & 0.324 & 0.450 & 0.004 & 0.0029 \\
5 & 40 & 250 & 0.350 & 0.500 & 0.007 & 0.0055 \\
9 & 30 & 70  & 0.326 & 0.350 & 0.032 & 0.0059 \\
9 & 30 & 250 & 0.252 & 0.300 & 0.015 & 0.0030 \\
9 & 40 & 250 & 0.250 & 0.525 & 0.021 & 0.0056 \\
\hline
\end{tabular}
\label{tab:Table_measuredsuppression_width_height}
\end{table}

In addition, to verify that the depths extracted with this model are physically consistent with the photon interaction range, we also calculated the photon-absorption distribution in germanium using energy-dependent attenuation coefficients from Xraylib. The calculation also includes transmission losses through the Be and Kapton entrance window and is shown in Table~\ref{tab:absorptiondepth}.
The resulting mean absorption depths (0.03–0.56~mm for 8–50~keV) are orders of magnitude greater than the fitted defect depths (1–6~$\mu$m), confirming that the observed suppression originates from shallow regions near the sensor's entrance window rather than from absorption limits or deep bulk trapping. A summary of used data points and the estimated defect height, width, and depth is tabulated using Table~\ref{tab:Table_measuredsuppression_width_height}.



\subsection{Dead Time evaluation with Direct beam}
\label{sec:beamline_deadtimetests}

During the beam time at the BM05 beamline, we evaluated the detector’s count rate performance in high-flux conditions ($\sim$2-3~Mcps) by exposing it to a direct monochromatic X-ray beam. Measurements were taken at two shaping time values: 1~$\mu$s (targeted for high rate operation) and 6~$\mu$s (currently used for optimal energy resolution). Three ASIC gain settings were tested to study their effect on the count rate behavior. The data below correspond to the pixel 3, which has shown no signs of defects up to an exposure of a 50~keV photon beam.

The input count rate (ICR) was varied by adjusting the beamline slits, while maintaining a constant beam spot size of approximately 35~$\mu$m on the sensor. The ICR was derived from a standard photodiode, while the output count rate (OCR) was measured from the digital pulse processor (DPP).

The relationship between input and output count rates (ICR vs. OCR) was established using the paralyzable dead time model (eq.~\ref{eq:paralyzable_model_eq}), as discussed already in detail in Section~\ref{sec:deadtimeandpileup}.
\begin{figure}[htb!]
\centering
\includegraphics[width=\textwidth, height= 8.8cm]{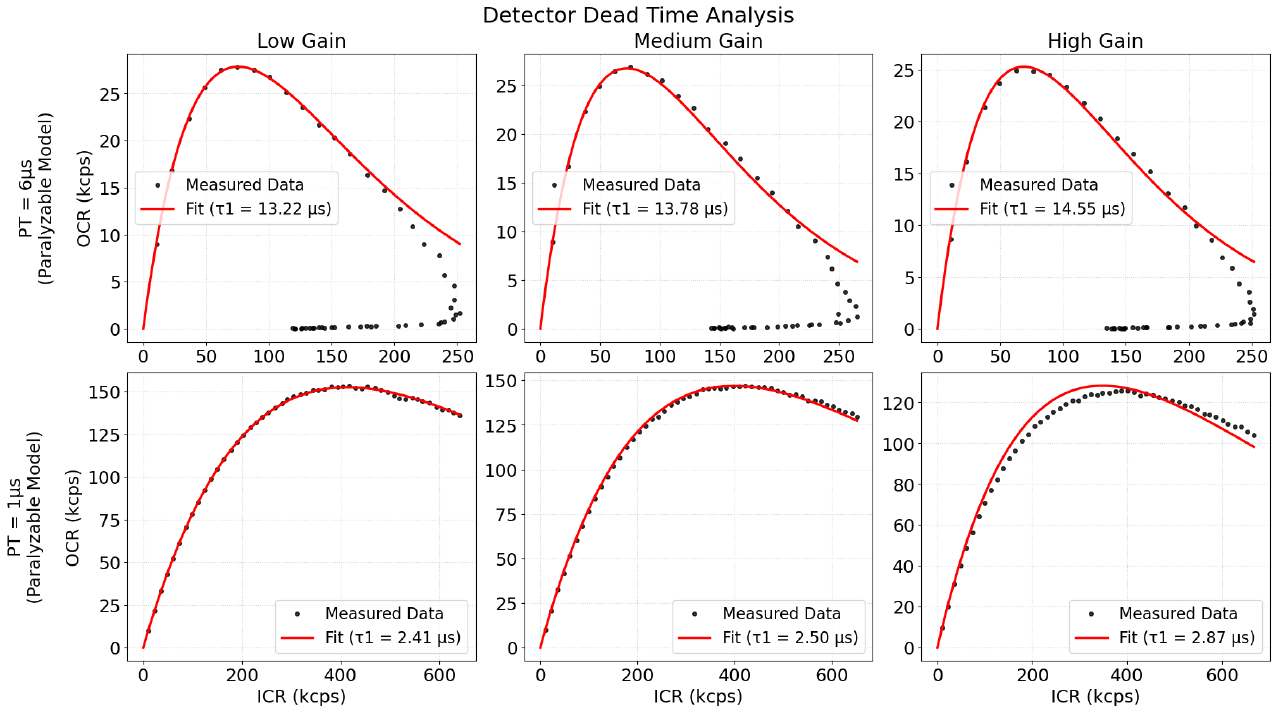}
\caption{ICR vs. OCR curves for shaping times of 1~$\mu$s and 6~$\mu$s across three ASIC gain settings. Red solid lines highlight the fit to measured data points.}
\label{fig:deadtime}
\end{figure}

We observe that at a PT value of 6~$\mu$s, the OCR saturates  at lower ICR values (30-40~kcps), delivering significant dead time losses, whereas, for 1~$\mu$s PT, the detector maintained linear response up in a broader range, and began saturating at ICR values above 150-200~kcps
The extracted $\tau_1$ values from the beamtime data agree well with the $\tau_1$ values (2.5 and 13~$\mu$s) at 1 and 6~$\mu$s PT, respectively, obtained from the lab fluorescence measurements in Section~\ref{sec:deadtimeandpileup}.


\section{Conclusion $\&$ Discussion}
The development of advanced detection systems is integral to meet the growing demands of next-generation synchrotron sources. In this work, we have discussed the new sensor design, laboratory characterization, and first on-beam tests of a new high-purity Germanium (HPGe) detector developed under the XAFS-DET work package of LEAPS-INNOV project. This detector focusses on high-resolution X-ray spectroscopy needs in the hard X-ray regime, where traditional silicon-based detectors reach their limits.

We have shown that the detector performs reliably across a wide energy range (6–50~keV) and remains stable under high input count rates. Future beam tests will explore performance at higher energies up to ~100 keV. Surface scans with a micro-focused beam confirmed excellent spatial uniformity in most regions, while a few defective areas, which we identified as shallow bulk defects later during the beam tests, showed reduced or no charge collection. A simple suppression model allowed us to estimate the depth and severity of these defects, confirming they are located close to the sensor surface and are likely linked to fabrication/handling issues. Future sensor production will include better quality control (ex, leakage current and capacitance checks) to identify such defects on an earlier stage.

The detector's performance improved significantly with the optimized electronics housing, particularly in reducing noise and enhancing energy resolution. For example, during recent tests at 5.89~keV (Mn-K$_{\alpha1}$), we observed a 30–40~eV improvement in energy resolution after modifications to the electronics box. Currently, the bottleneck is at the sensor level, as the best resolution we can achieve with an improved electronic configuration is $\sim$248~eV FWHM.

For future improvements, we plan to:
\begin{enumerate}
    \item 

Replace pogo-pin interconnections with indium bump bonding, which will reduce stray capacitance and improve energy resolution by minimizing noise contribution.
  \item 
Redesign the back-end electronics housing to isolate the DC-DC converter on the buffer board. Previous tests have shown that this isolation significantly reduces noise.
  \item 
Explore additional software-level corrections (e.g., digital charge-sharing rejection via advanced digital pulse processing, enabled by the Xpress4 system for enhanced performance in high-rate environments.
  \end{enumerate}
  \par
Overall, the prototype has shown strong potential for use in demanding synchrotron applications, and the lessons learned from this first version will guide the next steps in developing an improved HPGe detector system for synchrotron science.

\section*{Funding Resources}
This Project has received funding from the Horizon 2020 research and programme of the European Union under grant number 101004728.

\section*{Acknowledgement}
We thank the collaborators from ESRF, DESY, DLS, SOLEIL, MAXIV, PSI, Elettra, ALBA, INFN, EuXFEL, SOLARIS, XGlab, and Mirion for their immense support.\\
A sincere thanks is extended to the detector Group of ESRF $\&$ SOLEIL for their immense contribution in making all the mechanical advancements and detector Characterization.\\
We would also like to acknowledge the support $\&$ cooperation of the Optics Group of SOLEIL, especially Cyprian Wozniack and Lucie Depontailler, for allowing us access to the clean-room facilities for sensor modification and replacement for upcoming prototypes.\\
A very special thanks to Michele Bordessoule and Martin Chauvin for their valuable guidance and fruitful discussions.

\bibliographystyle{JHEP}
\bibliography{Leaps_Detector_Integration}
\end{document}